\documentclass[english]{article}
\usepackage[T1]{fontenc}
\usepackage[latin9]{inputenc}
\usepackage{color}
\usepackage{booktabs}
\usepackage{amsmath}
\usepackage{amssymb}
\usepackage{graphicx}
\usepackage{esint}

\makeatletter

\let\SF@@footnote\footnote
\def\footnote{\ifx\protect\@typeset@protect
    \expandafter\SF@@footnote
  \else
    \expandafter\SF@gobble@opt
  \fi
}
\expandafter\def\csname SF@gobble@opt \endcsname{\@ifnextchar[
  \SF@gobble@twobracket
  \@gobble
}
\edef\SF@gobble@opt{\noexpand\protect
  \expandafter\noexpand\csname SF@gobble@opt \endcsname}
\def\SF@gobble@twobracket[#1]#2{}
\providecommand{\tabularnewline}{\\}

\makeatother

\usepackage{babel}
\begin{document}

\title{Reconstructing Complex Dynamical Networks}

\title{Reconstructing Complex Networks From Dynamics}

\title{Reconstructing Networks from Dynamics}

\title{Reconstrucing Networks from Collective Dynamics -- An Introductory
Review}

\title{Finding Networks from Dynamics -- An Introductory Review}

\title{Revealing Networks from Dynamics \\
-- An Introduction}

\author{authors: \\
Marc Timme$^{1,2,3}$, Jose Casadiego$^{1}$}

\maketitle
$^{1}$Network Dynamics, Max Planck Institute for Dynamics and Self-Organization
(MPIDS), 37077 G\"{o}ttingen, Germany and 

$^{2}$Bernstein Center for Computational Neuroscience (BCCN) G\"{o}ttingen,
37077 G\"{o}ttingen, Germany

$^{3}$Institute for Nonlinear Dynamics, Faculty of Physics, University
of G\"{o}ttingen, 37077 G\"{o}ttingen, Germany \\
\\

Email: \textcolor{blue}{timme@in-cas.com} 
\begin{abstract}
What can we learn from the collective dynamics of a complex network
about its interaction topology? Taking the perspective from nonlinear
dynamics, we briefly review recent progress on how to infer structural
connectivity (direct interactions) from accessing the dynamics of
the units. Potential applications range from interaction networks
in physics, to chemical and metabolic reactions, protein and gene
regulatory networks as well as neural circuits in biology and electric
power grids or wireless sensor networks in engineering. Moreover,
we briefly mention some standard ways of inferring effective or functional
connectivity.\\
\\
Keywords: network reconstruction, network dynamics, network inference,
structural connectivity, functional connectivity, effective connectivity,
network topology, complex networks, synchrony \\
\\
PACS: 89.75.-k, 05.45.Xt, 87.18.-h, 87.16.Yc
\end{abstract}
\tableofcontents{}

\section{Where are you linked? }

\subsection{Relating connection topology of a network to its dynamics }

\textbf{\emph{Networks are everywhere.}} And most of them are dynamic.
From networks of biochemical reactions that regulate the metabolism
in the cells of our bodies to the neuronal circuits in our brains,
from social ties forming networks of our friendships and collaborators
to the power grids and internet that provide huge amounts of electric
energy and information every second. All of these systems form networks
of units that interact to yield complex, collective forms of functions
-- and all are crucial to our everyday life. 

\textbf{\emph{The interaction topology of complex networks strongly
impact their collective dynamics}} and thus the function of entire
systems. For many network dynamical systems, for instance in physics
and biology, the dynamics of individual units becomes more and more
accessible whereas their intricate web of interactions remains uncertain
or even often largely unknown. As an example, many constituents of
protein and gene interaction networks are well characterized but how
they interact and which pathways are relevant for suitable functioning
is not well understood \cite{Gardner:2003p9614,Palsson2006}. In neuroscience,
the number of units from which one can simultaneously measure neuronal
activity is increasing rapidly from a few units to hundreds of them
\cite{Brown:2004}. Still, identifying the synaptic connections of
a neuronal circuit by anatomical methods is mostly restricted to individual
synapses and computer-aided reconstruction based on optical methods
for more than two cells becomes available only since very recently
\cite{Briggman:2006,Helmstaedter:2011,Sporns:2005}. In social networks,
even in simple settings such as basic games, pairwise interactions
are roughly understood, but often both the (temporally varying) interaction
network and its collective consequences remain a riddle. Thus, reconstructing
the structure of interaction networks from (only) the collection of
local dynamical data constitutes a current open challenge, with applications
across the natural and social sciences as well as engineering.

\textbf{\emph{Network dynamics: forward vs. inverse problem.}} Yet,
the vast majority of research on network dynamics has focused on the
``forward direction'' of modeling and asked what types of collective
dynamics emerge from a network of given topology. Researchers from
the natural sciences and engineering systematically address the reverse
questions -- how to control a network or, more generally, how to design
networks for a desired dynamics and thus function and how to infer
topology from dynamics -- now at a rapidly increasing pace: In particular
in engineering, the design of systems for a specific function always
was core \cite{Maier:2009} and with the systems becoming more complex,
considering recurrently interacting networks becomes indispensable.
Conversely, complex systems in physics and biology require a view
on networked systems to understand how complex emergent phenomena
contribute to (possibly optimal) system's dynamics or function \cite{Prinz:2004,Rabinovich:2008,Timme:2007p14319,Memmesheimer:2006p10093,Srinivas:2011}.
Finally, also how one could perhaps redesign collective dynamics of
networks, e.g., gene and protein networks \cite{Francois:2004,Slusarczyk:2012}
poses challenges of frontier research. 

The inverse problem of how to infer interaction topology from network
dynamics constitutes the main topic of this review.

\subsection{Aims and options of network inference\label{sub:Aims-and-options}}

\vspace{0.3cm}

\begin{table}
\centering{}%
\begin{tabular}{cc}
\toprule 
\textbf{Property of connection} & \textbf{\emph{Distinctions}}\tabularnewline
 & \textbf{\small ... and examples thereof}\tabularnewline
\midrule
\midrule 
\multicolumn{1}{c}{structural vs. effective} & \multicolumn{1}{c}{\emph{direct interaction or statistical dependence?}}\tabularnewline
 & {\small EEG vs. connectomics for neural ``connectivity''}\tabularnewline
\midrule 
pure existence & \emph{presence or absence of links?}\tabularnewline
 & {\small does one gene directly influence another?}\tabularnewline
\midrule 
sign  & \emph{positive, negative, or mixed-sign interaction?}\tabularnewline
 & {\small phase-advancing or phase retarding (for coupled oscillators);}\tabularnewline
 & {\small activating or inhibiting (for gene and protein interactions);}\tabularnewline
\midrule 
directedness & \emph{directed or undirected (bi-directed) interactions?}\tabularnewline
 & {\small chemical vs. electrical neuronal synapses}\tabularnewline
\midrule 
type of interaction & \emph{continuous time or discrete time; linear vs. nonlinear?}\tabularnewline
 & {\small chemical vs. electrical synapses}\tabularnewline
 & {\small diffusive vs. nonlinear coupling}\tabularnewline
\midrule 
time scales & \emph{instantaneous, temporally localized or extended?}\tabularnewline
 & {\small slot communication (in mobile phones) vs. genetic interactions}\tabularnewline
\midrule 
spatial scales & \emph{local, global, non-local, bounded?}\tabularnewline
 & {\small message broadcasting, from wireless networks to cell tissue}\tabularnewline
\bottomrule
\end{tabular}\caption{\textbf{Levels of interest.} Which properties of the connections are
we interested in?\protect \\
\label{tab:Levels-of-interest.}}
\end{table}

\textbf{\emph{What do we aim to infer?}} Before addressing any inference
problem, we have to clarify what we actually want to find out about
the networked system and which level of detail we are interested in,
see Table \ref{tab:Levels-of-interest.}. For instance, we may want
to know \emph{effective} or \emph{functional connectivity} not caring
about individual interactions \emph{per se} but only about statistical
dependencies which the entire set of interactions yield between pairs
of units through the collective network dynamics. Inter-unit correlations
and various information-theoretic measures have been devised to solve
such problems. These often neglect the temporal dynamics as they use
temporal averages or statistical distributions of observables. Moreover,
effective connectivity may depend on the current collective state
and function of a given system and thus the same physical network
may display different effective topologies for different functions
or in different states. We may alternatively want to know \emph{structural
connectivity}, i.e. which unit directly interacts with which other
units (and how)? 

In this review, we focus on these direct physical interactions and
address various levels of detail. We also briefly present basic methods
to infer different types of effective connectivity . By construction,
such a review cannot be complete, also because the field is currently
developing at a breathtaking pace. We therefore select specific reconstruction
methods from those that are commonly used, appear promising for the
future of the field, or have been recently developed and form the
basis of current research.

\textbf{\emph{What can we learn about the connections}}\textbf{ }\textbf{\emph{of
a network from accessing the units' dynamics?}} Mathematically, inferring
the connectivity constitutes a high-dimensional inverse problem and
various methods have been devised to address this question. Every
inference method starts from different levels of pre-knowledge about
the system and has its own aims what to reconstruct, cf.~Table \ref{tab:Levels-of-interest.}.
We may be interested in whether the interactions are directed or undirected,
in whether or not a link exists, in the sign of the interaction dynamics,
the type of links (e.g., electric vs. chemical synapses, diffusive
vs. nonlinear coupling), the strengths and the temporal and spatial
scales of interactions etc.

\begin{figure}
\begin{centering}
\includegraphics[scale=0.06]{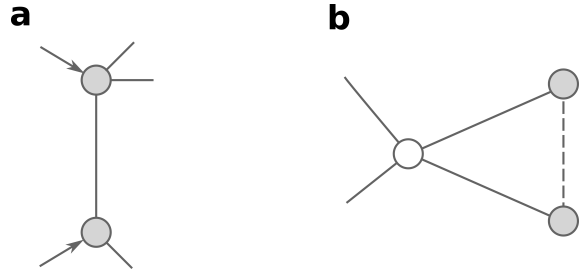}
\par\end{centering}

\caption{\textbf{Diverse dynamic impact of structural links on effective connectivity.}
(a) Structural links may not be detectable by certain correlation
measures due to strong independent driving signals (e.g. noise). For
example, strong inputs along two links (from within or outside the
network, marked by arrows) may decorrelate the dynamics of the two
nodes (gray disks) although they directly interact. (b) Common input
may create effective link without the structural link present. For
example, common input from a third node (open circle) may create effective
connectivity (dashed line) between two nodes (gray disks) that are
not directly connected. In both examples, (a) and (b), it may depend
on the entire collective state of the network (and the external inputs
it may receive) whether or not an effective connection is detected.
\label{fig:StucturalVSEffectiveLinks}}
\end{figure}

\textbf{Structural connectivity} may be very different from effective
or functional connectivity (Figure \ref{fig:StucturalVSEffectiveLinks}).
On the one hand, high correlations may exist between two units that
are not directly connected but only influenced by each other via a
third unit they both directly interact with. In general, such indirect
interactions may be induced not only by one third node, but equally
by the entire collective dynamics of a network. On the other hand,
even a strong direct interaction between two units does not necessarily
mean that their dynamics is highly correlated; correlations could
be submerged, e.g., by external noise or recurrent inputs the two
units receive, e.g., from two distinct other parts of the network,
or even from outside of it. 

In general, effective and structural connectivity are related in a
highly non-trivial way. In fact, a number of counter-intuitive phenomena
have been observed in various systems. For instance, recent work on
coupled oscillator networks highlight that under certain conditions
noise may aid to reconstruct structural connectivity from correlation-based
effective connectivity \cite{Ren:2010}; and one may detect small
world features in the functional connectivity even if it is derived
from randomly connected dynamical systems without any specific small-world
features \cite{Hlinka:2012}.

\textbf{\emph{Where do we start from?}}\textbf{ }It is important to
clarify which knowledge about the networked system we presuppose.
Is the collective dynamics known to be simple such as converging to
a fixed point of concentrations in biochemical networks or to a limit
cycle in coupled oscillator networks? Or do we expect more complex,
irregular, and perhaps unpredictable, chaotic or random types of dynamics?
Can we change the dynamics of units by interfering with the system
or can we just observe? Does the research question require a global
analysis in state space or do we focus on a specific dynamical state
where local analysis may suffice? We should answer these questions,
among others, before using or developing any inference method -- to
achieve reconstruction at the level we need with best quality and
minimal efforts, both experimentally and computationally.

Here we review recently developed approaches to inferring structural
connectivity of a network from accessing its collective dynamics.
The presented approaches assume various levels of pre-knowledge about
the system and may or may not require the observer to interfere with
the system. The article is structured as follows. We first clarify
in Section \ref{sec:Generic-Nonlinear-Dynamics} what we mean by a
network dynamical system, taking the view of continuous-time dynamics
described by coupled ordinary differential equations. In Sections
\ref{sec:Measuring-the-Response}, \ref{sec:Synchronizing-a-Model},
and \ref{sec:Direct-approaches}, we explain three principally different
classes of methods to obtain information about the structure of the
network topology from dynamical quantities. 

Section \ref{sec:Measuring-the-Response} illustrates basic theoretical
approaches based on measuring and evaluating the response of a network
to external perturbations or driving. As the response depends on both
the external driving signal and the interaction topology of the network,
sufficiently many driving-response experiments yield information about
the entire network topology. A second class of methods sets up a model
copy of the original system (Section \ref{sec:Synchronizing-a-Model})
and adapts the coupling matrix of the model so as to synchronize its
dynamics with the original. If synchronization is achieved, the topology
obtained for the model is taken as a proxy for the original. Finally,
a set of direct methods (Section \ref{sec:Direct-approaches}) rely
on measuring time series (or features thereof), evaluating temporal
derivatives, and exploiting smoothness assumptions to find solutions
to an optimization problem given by the restrictions by data.

We briefly comment on technical issues (Section \ref{sec:Technical-issues})
and mention basic core approaches for identifying effective connectivity
(Section \ref{sec:Correlation-based-methods}). These approaches rely
on simple linear correlation, maximum entropy principles and related
statistical inference methods. Finally, we provide an outlook (Section
\ref{sec:Open-Questions}) where we highlight current challenges,
point out aspects sometimes overlooked and show potential research
paths towards uncovering more of the topology of the networks that
surround us. 

This review on purpose is short but self-contained. It is intended
for researchers mainly in the physical and biological sciences and
engineering and assumes basic knowledge of dynamical systems and probability
theory. Let's start with the details.

\section{Nonlinear dynamics of networks\label{sec:Generic-Nonlinear-Dynamics}}

\subsection{Systems of ordinary differential equations describing networks}

Throughout the main part of this review, we consider networks of units
assumed to be described by systems of ordinary differential equations.
Discrete time maps coupled to a network are not discussed but typically
approaches similar to those presented here are viable in slightly
modified form. We discuss specific issues for systems of pulse-coupled
units, such as spiking neurons, in section \ref{sec:Direct-approaches}.
These are formally hybrid systems, i.e. mixtures of continuous-time
and discrete time systems. 

Assuming that interactions occur between pairs of coupled units, a
generic network dynamical systems is given by 

\begin{equation}
\frac{d}{dt}\boldsymbol{x}_{i}=\boldsymbol{f}_{i}\left(\boldsymbol{x}_{i}\right)+\sum_{j=1}^{N}J_{ij}\boldsymbol{g}_{ij}\left(\boldsymbol{x}_{i},\boldsymbol{x}_{j}\right)+\boldsymbol{I}_{i}(t)+\boldsymbol{\xi}_{i}(t)\label{eq:basicHighDimNetwork}
\end{equation}
where $i,j\in\{1,2,\ldots N\},$ $\boldsymbol{x}_{i}(t)=\left[x_{i}^{(1)}(t),x_{i}^{(2)}(t),\cdots,x_{i}^{(D)}(t)\right]^{\mathsf{T}}\in\mathbb{R}^{D}$
describes the state of the \emph{$i$}-th unit at time $t\in\mathbb{R}$,
and the functions $\boldsymbol{f}_{i}:\,\mathbb{R}^{D}\rightarrow\mathbb{R}^{D}$
and $\boldsymbol{g}_{ij}:\,\mathbb{R}^{D}\times\mathbb{R}^{D}\rightarrow\mathbb{R}^{D}$
mediate intrinsic and interaction dynamics of the \emph{D}-dimensional
units, respectively. The term $\boldsymbol{I}_{i}(t)$ represents
a vector of external driving signals (possibly random) and $\boldsymbol{\xi}_{i}(t)$
symbolically represents external noise. Finally, the $J_{ij}$ define
the interaction topology, in the simplest setting in terms of the
adjacency matrix $A$ such that $J_{ij}=A_{ij}=1$ if there is a direct
physical interaction from unit $j$ to $i$ and $J_{ij}=A_{ij}=0$
otherwise. In general, units' interaction may be higher order, e.g.
requiring terms like $\boldsymbol{h}_{ijk}\left(\boldsymbol{x}_{i},\boldsymbol{x}_{j},\boldsymbol{x}_{k}\right)$
added to the right hand side of (\ref{eq:basicHighDimNetwork}). For
instance in gene and protein interaction networks, a protein (the
so-called transcription factor, say unit $k$) is directly influencing
the rate of transcription of a gene (say, unit $j$) to a DNA segment
(say unit $i$). We do not treat such terms here explicitly. Their
relevance for network dynamical systems is discussed in\textbf{ }\cite{Timme:2014}. 

For some methods to infer effective connectivity, the functional form
of (\ref{eq:basicHighDimNetwork}) does not directly enter the inference
argument, other methods can be extended to include higher order terms
explicitly. We comment on higher order terms where appropriate (cf.
also Sect.~\ref{sec:Direct-approaches}).

\subsection{Rescaling, simplifications, and common interactions}

\textbf{\emph{Some a priori technical issues}}\textbf{:} Considering
(\ref{eq:basicHighDimNetwork}) as our basic level of description,
in case the dimension $D_{i}$ of the local dynamical system $i$
depends on unit $i$, we would just consider the maximal occurring
dimension $D=\max_{i\in\{1,\ldots,N\}}D_{i}$ and for each unit ignore
the $D-D_{i}$ dummy variables. This is done purely for notational
simplification. Note further that in general, the quantity $J_{ij}$
is a $D\times D$ matrix of coupling strengths $J_{ij}^{dd'}$, but
that for many paradigmatic model systems, only one of these elements
is non-zero, i.e. $J_{ij}^{dd'}=J_{ij}^{dd'}\delta_{d,d_{1}}\delta_{d',d_{2}}$such
that $J_{ij}=J_{ij}^{d_{1}d_{2}}$ is sufficient to describe the influence
of unit $j$ on $i$. 

Given a dynamical system of the form \eqref{eq:basicHighDimNetwork},
the right hand side is determined up to $N\times D$ additive constants
$C_{i}^{d}$ and one overall multiplicative constant $C_{0}$. Shifting
the state variables $x_{i}^{d}$ to co-moving frames and rescaling
time enables us to set $C_{i}^{d}=0$ for all $i$ and all $d$ and
$C_{0}=1$ without loss of generality.

For simplicity of presentation, we furthermore describe the methods
below as if they were for networks of coupled one-dimensional units
only. Often, different dimensions may be treated independently during
reconstruction.

We thus take the coupled equations
\begin{equation}
\boxed{\frac{d}{dt}x_{i}=f_{i}\left(x_{i}\right)+\sum_{j=1}^{N}J_{ij}g_{ij}\left(x_{i},x_{j}\right)+I_{i}(t)+\xi_{i}(t)}\label{eq:basicLowDimNetwork}
\end{equation}
as our basic network characterization we start from, where now the
variables $x_{i}$ and $x_{j}$ and the functions $f_{i}$ and $g_{ij}$
are treated as real scalars.

\textbf{\emph{Common Interaction Functions.}} Only \textbf{non-trivial
coupling} terms that are not identically zero, 
\begin{equation}
\frac{\partial g_{ij}}{\partial x_{j}}\neq0\label{eq:nontrivialCouplingFunction}
\end{equation}
for the relevant domain of arguments actually contribute to interactions,
so we assume all coupling functions $g_{ij}$ for which $J_{ij}\neq0$
to be non-trivial in this sense. A broad range of systems exhibits
\textbf{diffusive coupling} such that 
\begin{equation}
g_{ij}\left(x_{i},x_{j}\right)\propto\left.\left(x_{j}-x_{i}\right)\right.\label{eq:diffusiveCoupling}
\end{equation}
which is a special case of coupling functions 
\begin{equation}
g_{ij}\left(x_{i},x_{j}\right)=\left.\tilde{g}_{ij}\left(x_{j}-x_{i}\right)\right.\label{eq:phaseDifferenceCoupling}
\end{equation}
that depend on \textbf{state differences} (e.g. phase differences
for coupled oscillators) only. The simplest non-trivial form of interaction
is \textbf{linear coupling}, 
\begin{equation}
g_{ij}\left(x_{i},x_{j}\right)=A_{ij}x_{j}\,,\label{eq:directLinearCoupling}
\end{equation}
and does not depend on the dynamical variable of the unit it influences.

We have now set the stage to dive into specific inference approaches.

\section{Driving-response experiments\label{sec:Measuring-the-Response}}

\begin{figure}
\begin{centering}
\includegraphics[scale=0.1]{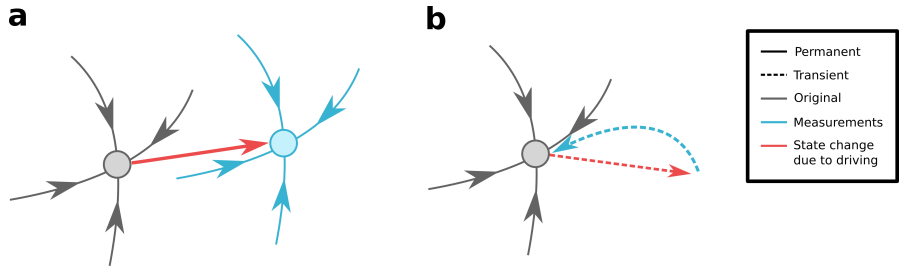}
\par\end{centering}

\caption{\textbf{Stable state approaches to network inference. }(a) Constant
external driving signal (parameter change) moves the stable fixed
point (gray) in state space to another position (blue). The difference
vector $v$ (red) depends on both the driving signal and the network
topology such that several measurements of $v$ under different driving
conditions yields information about the topology. (b) Same as (a)
but for transient perturbative signal: After the driving signal is
switched off, the dynamics relaxes back (dashed blue trajectory) towards
the original fixed point (gray). Now, the relaxation trajectory contains
information about the topology.\label{fig:Stable-state-approaches}}
\end{figure}

One idea of inferring network topology is to measure the collective
response of a network dynamical system to driving by external signals.
For instance, if a system exhibits a stable collective state (e.g.
fixed point or periodic orbit, cf.~Fig.~\ref{fig:Stable-state-approaches}),
it will relax back to that state after a transient input (pulse),
if the latter is not too strong (such as to not kick the system out
of the basin of attraction of the stable state) \cite{Timme:2002,Wiley:2006,Memmesheimer:2010,Jahnke:2008,Menck:2013}.
Here, the input signal (driving) effectively changes the initial condition
of the system, leaving the system features (given by the local and
interaction functions and their parameters) the same. Similarly, a
stable state of the system will generically move in state space (and
keep qualitatively the same stability properties) in response to sufficiently
weak, temporally constant external perturbations. These kinds of perturbations
effectively creates a non-identical but similar systems with different
parameters determined by the driving signal. 

Both the relaxation dynamics and the shift in state space in general
depend not only on the external signal (which unit is perturbed, how
and how strongly, i.e. known quantities), but also on the (unknown)
interaction topology of the network. Each collective response of the
system to an external perturbation yields a restriction on the network
topology such that sufficiently many driving-response experiments
may reveal the entire topology. In this section, we present the main
ideas underlying several related driving-response approaches. 

\subsection{Restrictions from local fixed point analysis\label{sub:Dynamics-Towards-FixedPoints}}

\textbf{} Recent efforts for developing methods to identify a network's
topology have emerged from the need to understand biological, in particular
gene regulatory networks \cite{Gardner:2003p9614,Tegner:2003p9754,diBernardo:2005p17589,Faith:2007p17716}.
Such networks consist of genes and proteins that interact with each
other within a cell \cite{Jong:2002,Po:2009}. These interactions
in particular control (indirectly) the rates at which genes are transcribed
into mRNA and  the regulatory features emerge via the interactions
and generally do not follow from the single-gene level \cite{Levine:2005}.
Gene regulatory networks and other reaction networks, e.g., in chemistry
or population dynamics, are often modeled by nonlinear differential
equations 
\begin{equation}
\dot{z}_{i}=f_{i}(z_{1},z_{2},\ldots,z_{N};\boldsymbol{p},\tilde{\boldsymbol{I}}),\label{eq:basicReactionNetwork}
\end{equation}
describing the rate of change of the expected numbers (or concentrations)
$z_{i}(t)$ of entities (e.g. genes, atoms and molecules or individual
organisms) at time $t$ in terms of their dependence on the number
of other entities \cite{Po:2009}.%
\footnote{From now on we write $\dot{z}$ for the rate of change $\frac{d}{dt}z$
of a variable $z$.%
} Here $\boldsymbol{p}$ is a set of parameters and $\tilde{\boldsymbol{I}}$
represents a set of external perturbations directed to the entities.
The $z_{i}$ are typically positive real numbers but mathematically
there is no restriction for them to also be negative. Under the assumption
that such a system is close to a steady state $\boldsymbol{z}^{*}=(z_{1}^{*},z_{2}^{*},\ldots,z_{N}^{*})$,
a stable fixed point where $f_{i}(z_{1},z_{2},\ldots,z_{N};\boldsymbol{p},\tilde{\boldsymbol{I}})=0$
for all $i$, the dynamics for perturbations $x_{i}(t)=z_{i}(t)-z_{i}^{*}$
from steady state concentrations of such nonlinear models may be approximated
to first order in the $x_{i}$ by 

\begin{equation}
\dot{x}_{i}=\sum_{j=1}^{N}J_{ij}x_{j}+I_{i}(t),\label{eq:dyntowfixed1}
\end{equation}
where the local Jacobian $J_{ij}=(\partial f_{i}/\partial z_{j})(\boldsymbol{z}^{*})$
is the effective interaction matrix given the steady state and $I_{i}(t)$
is assumed to be an external perturbation linearly coupling to deviations
of variables $x_{i}$. We remark that when deriving (\ref{eq:dyntowfixed1})
from (\ref{eq:basicReactionNetwork}) we implicitly use the relation
\begin{equation}
I_{i}(t)=\sum_{j}\tilde{I}_{j}(t)\left.\frac{\partial f_{i}}{\partial\tilde{I}_{j}}\right|_{\boldsymbol{z}=\boldsymbol{z}^{*},\tilde{\boldsymbol{I}}=\mathbf{0}}+\mathcal{O}\left(\tilde{I}_{i}\tilde{I}_{j}\right)\label{eq:firstOrderFPapproxDrivingIiIj}
\end{equation}
ignoring the second and higher order terms.

\subsubsection{Driving the system constantly to move a stable state\emph{ }\protect \\
\emph{(changing parameters)}}

The method presented now effectively moves the parameters and thus
in particular the fixed points of a system. Originally intended for
gene and protein interaction networks, Gardner and coworkers \cite{Gardner:2003p9614,Yeung:2002}
have explicated that reconstruction is possible, at least for small
networks. We here discuss two approaches of using the general form
(\ref{eq:dyntowfixed1}) to reconstruct the interaction network, i.e.
the $J_{ij}$. The first is based on driving the system (\ref{eq:dyntowfixed1})
by sufficiently small, temporally constant driving forces $I_{i}(t)=I_{i,m}$
to new fixed points $x_{i,m}^{*}\neq0$ close to the original one
$x_{i}^{*}=0$. As the original fixed point was structurally stable,
the new fixed point will generically exist and have qualitatively
the same stability properties if the $I_{i,m}$ are small enough (Fig.~\ref{fig:Stable-state-approaches}a).
This is because solutions of (\ref{eq:basicReactionNetwork}), in
particular fixed point solutions, typically vary continuously with
changing parameters (here: changing $I_{i,m}$ from zero) and there
is no bifurcation close to the parameters yielding a generic stable
fixed point $y_{i}^{*}$.  

The observed values assumed \emph{at} each new steady state together
with the (known) driving signals provide information about the interaction
topology $J_{ij}$. Performing several perturbation experiments $m\in\{1,\ldots,M\}$
yield $N\times M$ equations 

\begin{equation}
\sum_{j=1}^{N}J_{ij}x_{j,m}^{*}=-I_{i,m},\label{eq:dyntowfixed2}
\end{equation}
one for each experiment $m$ and for each unit $i$.

 After an arbitrary number $M$ of experiments, eq. (\ref{eq:dyntowfixed2})
in matrix form becomes

\begin{equation}
JX=Y,\label{eq:linearRestrictionsCORE}
\end{equation}
where $J\in\mathbb{R}^{N\times N}$ represents the connectivity among
the units, $X\in\mathbb{R}^{N\times M}$ the steady state values with
$X_{i,m}=x_{i,m}^{*}$, and $Y\in\mathbb{R}^{N\times M}$ the perturbations
$Y_{i,m}=-I_{i,m}$ that we assume to be known. This matrix equation
restricts the connectivity $J$ given the measured data $X$ and the
input perturbations $I$. The matrix equations constraining the full
network topology $J$ can be split into $N$ equations
\begin{equation}
\boldsymbol{J}_{i}X=\boldsymbol{Y}{}_{i}\,,\label{eq:linearRestrictionsCORESingleInputs}
\end{equation}
one for each input connectivity $\boldsymbol{J}_{i}:=(J_{i,1},\ldots,J_{i,N})\in\mathbb{R}^{1\times N}$
of a unit $i$. Thus, the same set of data $X$ restrict all the sets
of units providing interactions to $i\in\{1,\ldots,N\}$ but the data
$\boldsymbol{Y}{}_{i}=(Y_{i,1},\ldots,Y_{i,M})^{\textsf{T}}$ are
unit dependent. This reduction to $N$ individual equations also admits
to split the computational effort for solving them. The problem becomes
trivially parallelizable because for different $i$, these restrictions
(\ref{eq:linearRestrictionsCORESingleInputs}) are independent in
the sense that reconstruction of the input coupling strengths to each
unit $i$ can be performed without taking care of input coupling strengths
of other units $k\neq i$.

\subsubsection{Observe relaxation to stable state after transient driving \emph{}\protect \\
\emph{(changing initial conditions)}}

A second approach assumes that the quantities $y_{i}$ (and thus the
$x_{i}$) are perturbed such that at time $t_{0}$ we have $x_{i}(t_{0})=x_{i}^{(0)}$
and the transient dynamics $y_{i}(t)$ of relaxation back to the original
fixed point $y_{i}^{*}$ (and thus $x_{i}^{*}=0$) are observed at
a sequence of times $t_{m}>t_{0}$, $m\in\{1,\ldots,M\}$. This yields
the same type of equation (\ref{eq:linearRestrictionsCORE}), but
now with the $Y_{i,m}=-\hat{\dot{x}}_{i,m}$ being estimated of the
derivatives $\dot{x}{}_{i}(t_{m})$. We remark that these derivatives
may be estimates in various ways, each of them requiring a resolution
of the measured data on sufficiently small time scales, cf. section
\ref{sub:Reconstruction-by-observation}.\textbf{ }In this second
approach, the different times the transient dynamics is measured replaces
the different driving experiments in the first approach. Finally,
both approaches can of course be combined, several experiments evaluated
at several time points, again yielding the same form of restrictions
(\ref{eq:linearRestrictionsCORE}).

\subsection{Solving the restricting equations\label{sub:Solving-the-restricting-equations}}

How can we finally obtain the coupling elements $J_{ij}$ and thus
the interaction network? In principle, solving the matrix equation
(\ref{eq:linearRestrictionsCORE}) yields the interaction matrix $J$
as a function of the known data $X$ and $Y$. One may naively assume
that it is directly solvable once the number of experiments equals
the number of units in the network, $M=N$. However, this problem
can be numerically ill-conditioned \cite{Stoer:1993} for large $N$,
such that the result is not reliable. In addition, as also stressed
in \cite{Gardner:2003p9614}, the results may be sensitive to noise
in the measured data.

A way to overcome this problem is by performing (many) more experiments
than nodes available, $M\gg N$, thus \emph{over-determining} (\ref{eq:linearRestrictionsCORE}).
In general, due to noise and measurement inaccuracies, this yields
the system (\ref{eq:linearRestrictionsCORESingleInputs}) to be inconsistent
such that there is no vector $\boldsymbol{J}_{i}$ that satisfies
all constraints. It will still be possible to find a robust approximation
$\boldsymbol{\hat{J}}_{i}$ that minimizes the error between the predicted
dynamics $\boldsymbol{J}_{i}X$ and the actual dynamics $\boldsymbol{Y}{}_{i}$
for a given node. Specifically, this error function may be modeled
as

\begin{equation}
E_{i}(\boldsymbol{\hat{J}}_{i})=d(\boldsymbol{Y}_{i},\,\boldsymbol{\hat{J}}{}_{i}X),\label{eq:minimization}
\end{equation}
where the distance measure $d(\boldsymbol{v},\boldsymbol{w})=\|\boldsymbol{v}-\boldsymbol{w}\|_{p}^{p}$
between two vectors $\boldsymbol{v},\boldsymbol{w}\in\mathbb{R}^{M}$
is commonly defined in terms of the $p$th power 
\begin{equation}
d(\boldsymbol{v},\boldsymbol{0})=\|\boldsymbol{v}\|_{p}^{p}=\left(\sum_{m=1}^{M}v_{m}^{p}\right)\label{eq:pNorm^p}
\end{equation}
of an $L_{p}$-norm with $p\geq1$, due to its convexity properties.
This guarantees that any local minimum of $E_{i}$ is also a global
minimum \cite{Boyd:2009}. Particularly, the $L_{2}$-minimization
criterion,

\begin{equation}
E_{i}(\boldsymbol{\hat{J}}_{i})=\|\boldsymbol{Y}_{i}-\boldsymbol{\hat{J}}{}_{i}X\|_{2}^{2},\label{eq:l2minimization}
\end{equation}
is of great importance because it has an analytical solution for its
extremum. Equating to zero the derivatives of the error function with
respect to the matrix elements, $\frac{\partial}{\partial J_{ik}}E_{i}\left(\boldsymbol{\hat{J}}{}_{i}\right)\stackrel{!}{=}0$,
yields an analytical solution (see Appendix \ref{sec:l2minimization})
to $L_{2}$ error-minimization given by

\begin{equation}
\boldsymbol{\hat{J}}{}_{i}=\boldsymbol{Y}{}_{i}X^{\mathsf{T}}(XX^{\mathsf{T}})^{-1}.\label{eq:l2minimization2}
\end{equation}

Evaluating such equations for all $i\in\{1,\ldots,N\}$ yields the
complete reconstructed network $\hat{J}$. This mathematical form
of minimum $L_{2}$-norm solution is implemented in many mathematical
packages (e.g. as the \texttt{\textcolor{black}{mrdivide}} function
in Matlab~\cite{matlab:2006} or the \texttt{LeastSquares} function
in Mathematica \cite{Wolfram:2010}). We explicate that the obtained
off-diagonal terms $\hat{J}_{ij}$ serve as the best estimate (in
the procedural sense using the $L_{2}$-minimization above) for the
coupling constants $J_{ij}$ ; at the same time, the diagonal elements
$J_{ii}$ are not relevant for the network topology because the influence
of these terms on the dynamics of unit $i$ is physically indistinguishable
from an intrinsic drive to $i$ included in the local dynamics specified
by $f(x_{i})$ in (\ref{eq:basicLowDimNetwork}), cf. eq.~(\ref{eq:nontrivialCouplingFunction}).

Experimentally, it is in principle possible to over-determine a system
of equations by performing repeated measurements on the network until
the condition $M\gg N$ is achieved. Nevertheless, it may often seem
unsuitable for large networks due to the large number of experiments
that would be required. 

So, if the size of the network is an issue or the number of available
measurements insufficient to over-determine the system, we have $M<N$.
Assuming that the network is sparse (i.e. that each unit is connected
with a small number of others and thus many connection strengths are
$J_{ij}=0$), may still yield the collection of all network links.
This implies that several $J_{ij}$ are effectively set to zero, therefore
decreasing the amount of unknown coefficients to be solved for. It
leaves us with the problem of finding which links are actually present
and which are not. We present two related options to do so.

\subsubsection{Sparse solution with a bounded connectivity per unit\emph{ }}

If an upper bound for the number of links $K_{i}<M$ is known, we
may assume only $M<N$ experiments are available and we have a rough
idea of how many nodes (at most) are connected to a particular node.
In particular, assume that the number of incoming connections for
node $i$ is given by at most $K_{i}<M$ \cite{Gardner:2003p9614}.
This assumption shifts the system (\ref{eq:linearRestrictionsCORESingleInputs})
from having more unknowns than constraints, $M<N$, to have more constraints
than unknowns, \emph{$M>K_{i}$}, therefore, implicitly over-determining
the system. It means that out of the $N$ nodes present in the network
only $K_{i}$ of them are chosen to be part of the system of equations
for node $i$. Such assumption may be done when there is some \emph{a
priori }information about the network's connectivity and dynamics.

Specifically, the system of equations (\ref{eq:linearRestrictionsCORESingleInputs})
may in principle be rewritten as

\begin{equation}
\boldsymbol{B}_{i}Z_{i}=\boldsymbol{Y}{}_{i},\label{eq:maximumconnectivity}
\end{equation}
where $\boldsymbol{B}_{i}\in\mathbb{R}^{1\times K_{i}}$ is the reduced
connectivity vector for node \emph{$i$} that contains the coupling
strengths for the selected nodes and $\boldsymbol{Z}_{i}\in\mathbb{R}^{K_{i}\times M}$
is a matrix that contains the states of such nodes. If we knew which
$K_{i}$ of the $N-1$ possible connections actually contributed,
we could use eq.~(\ref{eq:l2minimization2}) to solve (\ref{eq:maximumconnectivity})
using $L_{2}$-minimization yielding

\begin{equation}
\hat{\boldsymbol{B}}_{i}=\boldsymbol{Y}{}_{i}Z_{i}^{T}(Z_{i}Z_{i}{}^{T})^{-1},\label{eq:dyntowfixed6}
\end{equation}
where $\hat{\boldsymbol{B}}_{i}$ is the best approximation to $\boldsymbol{B}_{i}$.

Yet, it so far remains unclear which of the $\binom{N}{K_{i}}=\frac{N!}{K_{i}!(N-K_{i})!}$
possible combinations of incoming connections is best suited for reproducing
the dynamics of $i$. In principle, $\hat{\boldsymbol{B}}_{i}$ may
be calculated for each combination of $K_{i}$ genes, and the combination
that yields the smallest value of the $L_{2}$ norm $\left\Vert \hat{\boldsymbol{B}}_{i}\right\Vert _{2}$
in (\ref{eq:dyntowfixed6}) may be chosen as the best estimate for
$J_{i}$. The efficiency of such a procedure relies in number $K_{i}$
of interactions per node. Hence, choosing a proper $K_{i}$ aiming
to recover the largest number of real interactions with the smallest
number of false positives is a key factor to achieve a successful
topological reconstruction.

\subsubsection{Maximizing the sparseness of the connectivity matrix\label{sub:Maximizing-the-sparseness}}

If\emph{ }$M<N$ and the\emph{ }$K_{i}$ are unknown, cannot be estimated
or there are too many of them (making the combinatorial search practically
impossible) maximizing the number of zero entries in $\boldsymbol{J}$,
(i.e. minimizing the $K_{i}$ and thus maximizing sparseness) may
be a way to solve (\ref{eq:linearRestrictionsCORESingleInputs}).
This approach is particularly useful if the only \emph{a priori }knowledge
about the network's connectivity is some sparsity.

For general matrix equations
\begin{equation}
A\boldsymbol{y}=\boldsymbol{b},\label{eq:SVD1maintext}
\end{equation}
where $A\in\mathbb{R}^{m\times n}$, $\boldsymbol{y}\in\mathbb{R}^{n\times1}$
and $\boldsymbol{b}\in\mathbb{R}^{m\times1}$, singular value decomposition
(SVD) of $A$ according to 
\begin{equation}
A=U\Sigma V{}^{\mathsf{T}},\label{eq:SVD2maintext}
\end{equation}

yields an analytic solution 
\begin{equation}
\boldsymbol{y}=V\tilde{\Sigma}U{}^{\mathsf{T}}\boldsymbol{b}+V\boldsymbol{c},\label{eq:SVD3finalmaintext}
\end{equation}

where $\tilde{\Sigma}=\Sigma^{\mathsf{T}}\left(\Sigma\Sigma^{\mathsf{T}}\right)^{-1}$that
parametrizes the space of all solutions through the vector $\boldsymbol{c}\in\mathbb{R}^{n\times1}$
with $c_{i}=0$ for $i\in\{1,...,r\}$ and $r=Rank(\boldsymbol{A})$. 

In our reconstruction problem, we are seeking to maximize the number
of zero entries in $\boldsymbol{J}$ based on solving the restricting
equations (\ref{eq:linearRestrictionsCORESingleInputs}) for $\boldsymbol{J}{}_{i}$
as we solved (\ref{eq:SVD1maintext}) for $\boldsymbol{y}$. Consider
the transpose

\begin{equation}
X^{\mathsf{T}}\boldsymbol{J}_{i}^{\mathsf{T}}=Y{}_{i}^{\mathsf{T}}\label{eq:maximizingSparness1}
\end{equation}
of (\ref{eq:linearRestrictionsCORESingleInputs}). The analogous SVD-based
solution then reads

\begin{equation}
\boldsymbol{J}{}_{i}^{\mathsf{T}}=V\tilde{\Sigma}U{}^{\mathsf{T}}\boldsymbol{Y}{}_{i}^{\mathsf{T}}+V\boldsymbol{c},\label{eq:maximizingSparseness2}
\end{equation}
where $U\in\mathbb{R}^{M\times M}$, $V\in\mathbb{R}^{N\times N}$,
$\tilde{\Sigma}\in\mathbb{R}^{N\times M}$ and $\boldsymbol{c}\in\mathbb{R}^{N\times1}$
is a vector of remaining coefficients parametrizing the solution space.
Thus, the set of all possible solutions for $\boldsymbol{J}{}_{i}$
is given by (\ref{eq:maximizingSparseness2}). The goal now is to
pick the sparsest solution from this set. Therefore, eq. (\ref{eq:maximizingSparseness2})
may be posed as the overdetermined ($M>N-r$) problem

\begin{equation}
V\tilde{\Sigma}U^{T}\boldsymbol{Y}_{i}^{T}=-V\boldsymbol{c}.\label{eq:maximizingSparseness3}
\end{equation}
Minimizing the $L_{1}$ error

\begin{equation}
E_{i}(\boldsymbol{c})=\|V\tilde{\Sigma}U^{T}\boldsymbol{Y}_{i}^{T}+V\boldsymbol{c}\|_{1},\label{eq:maximizingSparseness4}
\end{equation}
yields a sparse solution \cite{Boyd:2009}. However, unlike the $L_{2}$
minimization, $L_{1}$ minimization has no analytical solution, so
choosing an appropriate iterative algorithm to solve it is essential.
The Barrodale Roberts algorithm \cite{Barrowdale:1974} provides a
particularly fast solver that has been vastly used in the field of
network reconstruction \cite{Timme:2007p14319,Srinivas:2011,Yeung:2002,VanBussel:2011}.

\emph{Remarks.} The core equations (\ref{eq:DEshortform}) also provide
the option to reconstruct network connectivity via maximizing sparseness
of the network and there is a particular relation to what is known
as compressive sensing cf., e.g.~\cite{Wang:2011} . 

In general, linearization of dynamical equations, e.g. linearizing
in state variables close to fixed points, often well approximates
nonlinear dynamics\textcolor{red}{. }\textcolor{black}{This seems
to hold for gene regulatory networks \cite{Yeung:2002} as well as}
in models of Drosophila segmentation networks \cite{Dassow:2000}\textcolor{red}{{}
}and may thus be of general use across systems. For gene and protein
interaction networks, often single genes are selected for perturbations
in an experiment, with the danger of providing non-generic restrictions
in (\ref{eq:linearRestrictionsCORESingleInputs}). Finally, for some
systems increasing the number of experiments may reduce the resulting
computational costs such that this trade-in may be considered.

\subsection{Driving the system's state to a fixed point\label{sub:Driving-the-system's}}

One may also infer network structure by externally driving the system
to a fixed point and shifting component values of the fixed point
for individual units \cite{Yu:2010a,Yu:2010b}. As before, the differences
between pairs of steady-state responses are analyzed. Let us describe
the network as

\begin{equation}
\dot{x}_{i}=f_{i}(x_{i})+\sum_{j=1}^{N}A_{ij}g_{ij}(x_{i},x_{j})+I_{i},\label{eq:MovedFix0}
\end{equation}
where the $A_{ij}\in\{0,1\}$ are the entries of the adjacency matrix
specifying only if an interaction from $j$ to $i$ is present ($A_{ij}=1$)
or not ($A_{ij}=0$), the $g_{ij}(x_{i},x_{j})$ are the coupling
functions from $j\in\left\{ 1,2,\ldots,N\right\} $ to $i$, and $I_{i}$
is the driving signal applied to unit $i$. It was demonstrated by
Yu and Parlitz \cite{Yu:2010a} that under driving signals 
\begin{equation}
I_{i}=-\left(x_{i}-\hat{x}_{i}\right)\theta,\label{eq:MovedFix2-1}
\end{equation}
with sufficiently large gain factor $\theta\in\mathbb{R}$ and Lipschitz
continuous $f_{i}$ and $g_{ij}$ , the network may be driven to a
globally stable fixed point $\boldsymbol{x}^{*}:=(x_{1}^{*},x_{2}^{*},\ldots,x_{N}^{*})^{\mathsf{T}}\in\mathbb{R^{N}}$
that is arbitrarily close to a predetermined point $\hat{\boldsymbol{x}}:=(\hat{x}_{1},\hat{x}_{2},\ldots,\hat{x}_{N})^{\mathsf{T}}\in\mathbb{R^{N}}$,
independent of the initial conditions \cite{Yu:2010a}\textcolor{green}{{}
}. At such fixed point we have 

\begin{equation}
(x_{i}^{*}-\hat{x}_{i})\theta=f_{i}\left(x_{i}^{*}\right)+\sum_{j=1}^{N}A_{ij}g_{ij}\left(x_{i}^{*},x_{j}^{*}\right)\label{eq:MovedFix1}
\end{equation}
for all $i$. To understand how the network responds to changes in
$\hat{\boldsymbol{x}}$, let us define 

\begin{equation}
\Delta_{i}:=f_{i}(\hat{x}_{i})+\sum_{j=1}^{N}J_{ij}g_{ij}\left(\hat{x}_{i},\hat{x}_{j}\right)-\left[f\left(x_{i}^{*}\right)+\sum_{j=1}^{N}J_{ij}g_{ij}\left(x_{i}^{*},x_{j}^{*}\right)\right],\label{eq:MovedFix4}
\end{equation}
hence, eq. (\ref{eq:MovedFix1}) may be rewritten in terms of $\Delta_{i}$
as 

\begin{equation}
\left(x_{i}^{*}-\hat{x}_{i}\right)\theta=f_{i}(\hat{x}_{i})+\sum_{j=1}^{N}A_{ij}g_{ij}\left(\hat{x}_{i},\hat{x}_{j}\right)-\Delta_{i}.\label{eq:MovedFix5}
\end{equation}

The main idea at this point is to check whether unit $k$ couples
to unit $i$ by evaluating how $x_{i}^{*}$ reacts to the shifting
of $x_{k}^{*}$ through $\hat{x}_{k}$. Therefore, let us set

\begin{equation}
\hat{x}_{j}=\begin{cases}
\hat{x}_{k} & \mbox{if }j=k\\
0 & \mbox{if }j\neq k
\end{cases},\label{eq:MovedFix6}
\end{equation}
and evaluate (\ref{eq:MovedFix5}) at this point, yielding 
\begin{equation}
x_{i}^{*}\theta=A_{ik}g_{ik}\left(0,\hat{x}_{k}\right)-\Delta_{ik}+f_{i}(0)+\sum_{j\neq k}^{N}A_{ij}g_{ij}\left(0,0\right).\label{eq:MovedFix7}
\end{equation}

Shifting the same component twice to $\hat{x}_{k}^{(1)}$ and $\hat{x}_{k}^{(2)}$,
resp., fixes a reference frame and thereby yields equations characterizing
the difference between responses of a given unit $i$ to a shifted
fixed point component for unit $k$. It results in 

\begin{equation}
\left[x_{i,2}^{*}-x_{i,1}^{*}\right]\theta=A_{ik}\left[g_{ik}\left(0,\hat{x}_{k,2}\right)-g_{ik}\left(0,\hat{x}_{k,1}\right)\right]+[\Delta_{ik,1}-\Delta_{ik,2}],\label{eq:MovedFix8}
\end{equation}
a condition that may be rewritten as

\begin{equation}
S_{ik}\theta=A_{ik}\eta_{ik}+\lambda_{ik}.\label{eq:MovedFix9}
\end{equation}
We remark that the differences $[\Delta_{ik,1}-\Delta_{ik,2}]$ in
(\ref{eq:MovedFix8}) are not known but the general form (\ref{eq:MovedFix9})
may be used to reveal whether they are zero, $\lambda_{ik}=0$, or
not: Given that we are dealing with entries of the adjacency matrix,
we may infer two possible outcomes from (\ref{eq:MovedFix9}), whether
system $k$ is coupled to $i$ or not. Specifically, 
\begin{equation}
S_{ik}\theta=\begin{cases}
\eta_{ik}+\lambda_{ik} & \text{if unit \mbox{\emph{k} coupled to \emph{i} }}\\
\lambda_{ik} & \text{if not}
\end{cases}\label{eq:MovedFix10}
\end{equation}

It was also demonstrated by Yu and Parlitz \cite{Yu:2010a} that $\Delta_{i}$
decreases with $\theta$ if $f_{i}$ and $g_{ij}$ are Lipschitz continuous.
This permits to discriminate whether there is a coupling between a
pair $k\rightarrow i$. Especially, when $\theta$ is sufficiently
large, the $|S_{ik}\theta|$ values may be classified into sets $\mathbb{I}_{0}$
and $\mathbb{I}_{1}$, non-coupled and coupled sets, respectively.
To construct such sets, Yu and Parlitz \cite{Yu:2010a} propose to:
\begin{itemize}
\item For fixed $k,$ organize the $|S_{ik}\theta|$ values in an ascending
order, i. e., the values should be arranged into a new series $z$
where $z_{k,j}<z_{k,j+1}<\ldots$ that defines the indexing $j$ of
the $z_{k,j}$'s.
\item Establish the critical values of each set. In this case, the critical
values $j_{c}$ and $j_{c+1}$ define the end and the beginning of
$\mathbb{I}_{0}$ and $\mathbb{I}_{1}$, respectively. Yu and Parlitz
suggest to find $j_{c}$ by requiring the distance between any element
from $\mathbb{I}_{1}$ with respect to $z_{i,1}$ to be larger than
twice the size of $\mathbb{I}_{0}$, $z_{k,j}-z_{k,1}\geq2(z_{k,j_{c}}-z_{k,1})$
for all $j>j_{c}$.
\end{itemize}
Finally, by performing this process on every unit, the topology of
the network may be reconstructed. 

\emph{Remarks: }The approach relies on the feasibility of (i) perturbing
the systems in a specific manner, and (ii) measuring the steady states,
suggesting that it is model independent to a large extent in that
it does not in principle require knowledge of local dynamics or coupling
functions (yet it requires these to be Lipschitz continuous). These
features may make the approach of interest under certain conditions
where only little pre-knowledge about the system is available. Yet
the approach requires substantial control over the system, in particular,
the option of externally driving every unit (independently) constitutes
a major requirement. The study \cite{Yu:2010a} does not state how
indirect actions are treated, for instance, how is an indirect effect
from unit $k$ via $k'$ onto $i$ distinguished from direct interactions
from $k$ to $i$? Possibly, driving $k$ may indirectly affect $i$
only weakly and this potentially second order contribution could be
treated in a perturbative way.

\subsection{Distributed perturbations to collective periodic dynamics }

The approaches presented above (sections \ref{sub:Dynamics-Towards-FixedPoints}-\ref{sub:Driving-the-system's})
required the existence of fixed points either in the original system
or in the presence of sufficiently strong external driving. Yet, more
complex dynamics prevails in a large range of biological, physical
or artificial systems. The second most simple invariant dynamics are
periodic orbits and often arise as limit cycles of coupled oscillatory
units, thus asking for a generalization beyond simple fixed point
approaches. Even more complex dynamics, e.g. collective chaos, is
treated by a direct approach below in section \ref{sub:Reconstruction-by-observation}.

Is it possible to infer network topology from driving-response experiments
also for oscillator networks? Below we positively answer this question,
at the same time showing that distributed driving signals not precisely
targeted to one or a few units are at least equally appropriate to
infer network topology. Several theoretical model studies of coupled
oscillators \cite{Kori:2004p10015,Kawamura:2008p14622,Zanette:2004p9940,Radicchi:2006,Timme:2006p987,Arenas:2006p10049}
have shown that the response of single units in a network to constant
or periodic driving signals as well as the transient dynamics of synchronization
depend on the network topology. Some recent works \cite{Radicchi:2006,Timme:2006p987}
helped us to understand specific quantitative influence of structural
features on the response and how the network response provides some
information about the structure (and the driving signal). For instance,
the magnitude of responses seem to decay exponentially with distance
from the driving node \cite{Radicchi:2006}, and the coarse-scale
connectivity among connected components may qualitatively determine
to which degree network dynamics is coordinated globally \cite{Timme:2006p987}.
Further developing such insights, a follow-up work \cite{Timme:2007p14319}
presents a method of reconstructing network topology from systematic
measurements of network responses to temporally constant, distributed
driving signals in coupled phase-oscillator networks. 

\begin{figure}
\begin{centering}
\includegraphics[scale=0.08]{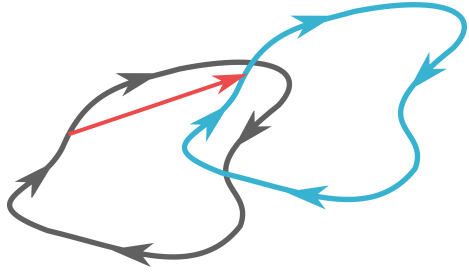}\caption{\textbf{Topology revealed by driving a stable periodic orbit:} As
for fixed point approaches discussed above (Section~\ref{sub:Dynamics-Towards-FixedPoints}
and Fig.~\ref{fig:Stable-state-approaches}), the difference vector
$v$ (red) depends on both the driving signal and the network topology
such that several measurements of $v$ under different driving conditions
yields information about the topology.\textbf{ \label{fig:Topology-revealed-by}}}

\par\end{centering}

\end{figure}

The basic idea is that any network displaying a stable invariant dynamics,
not just fixed points, yield a specific response to a given perturbation
as a consequence of the network's topology and the perturbation itself
\cite{Kori:2004p10015,Zanette:2004p9940,Timme:2006p987} cf. Fig.
\ref{fig:Topology-revealed-by}. If the perturbations are small, the
invariant set is typically \emph{qualitatively} unchanged and only
slightly moved in state space. Keeping track of which driving signals
resulted in which responses, we can collect evidence about the interactions
among units in a network. Sufficiently many repetitions of appropriate
driving-response experiments then yield the network's topology.

Weakly coupled limit cycle oscillators are well-characterized by ignoring
(in the long time limit) amplitude responses to coupling and by modeling
them as phase-oscillators with coupling via their phase-differences
only. A method to infer network topology for coupled phase oscillators
with arbitrary stable, phase-locked dynamics has been presented in
\cite{Timme:2007p14319}. One key observation is that the phase differences
(yet not the phases themselves) in such systems converge with time
and that comparing differences of phase differences among different
driving conditions yield restrictions to network topology. The network
dynamics is given by 

\begin{equation}
\dot{\phi}_{i}=\omega_{i}+\sum_{j=1}^{N}J_{ij}g_{ij}(\phi_{j}-\phi_{i})+I_{i,m},\label{eq:DistPert1}
\end{equation}
where $\phi_{i}(t)$ and $\omega_{i}$ are the phase and natural frequency
of oscillator $i$, respectively, $J_{ij}$ the connection strength
from oscillator $j$ to $i$ and $I_{i,m}$ is a temporally constant
driving signal applied to $i$ during the experiment $m$. We assume
that in the absence of driving, $I_{i,m}\equiv0$, the network is
in a phase-locked state where $\dot{\phi}_{j}-\dot{\phi}_{i}=0$ for
all $i,j$. We remark that one, several or all units may be perturbed
during each given experiment, such that driving can be arbitrarily
distributed and effectively changes the frequencies of the driven
oscillators. As for the approaches relying on fixed points (section
\ref{sub:Dynamics-Towards-FixedPoints}), the existence of a stable
periodic orbit (and thus in particular a phase-locked state) implies
that sufficiently small constant perturbations yield a (only slightly
moved and slightly different) stable periodic orbit.

If for a given driving condition $m$, the dynamics becomes phase-locked,
the phase differences 
\begin{equation}
\Delta_{ij,m}(t)=\phi_{j,m}(t)-\phi_{i,m}(t)\label{eq:PhaseDifferences}
\end{equation}
become constant in time, $\Delta_{ij,m}(t)\rightarrow\Delta_{ij,m}^{*}:=\lim_{t\rightarrow\infty}\left(\phi_{j,m}(t)-\phi_{i,m}(t)\right)$
because all oscillators move at the same collective frequency

\begin{equation}
\Omega_{m}=\omega_{i}+\sum_{j=1}^{N}J_{ij}g_{ij}(\phi_{j,m}-\phi_{i,m})+I_{i,m}.\label{eq:DistPert2}
\end{equation}
Hence, if the network is perturbed by a sufficiently small driving
signal, the original phase-locked state (for $I_{i,m}\equiv0$) is
slightly moved such that $|\Delta_{ij,m}^{*}-\Delta_{ij,0}^{*}|\ll1$
and there is a small difference between the perturbed and non-perturbed
collective frequencies $\Omega_{m}$ and $\Omega_{0}$. Defining the
effective frequency difference $D_{i,m}:=\Omega_{m}-\Omega_{0}-I_{i,m}$
of oscillator $i$, and approximating the arbitrarily nonlinear coupling
functions $g_{ij}$ by a first order Taylor expansion around $\Delta_{ij,0}^{*}$
we obtain

\begin{equation}
D_{i,m}=\sum_{j=1}^{N}\hat{J}_{ij}\theta_{j,m}\label{eq:DistPert3}
\end{equation}
where $\theta_{j,m}:=\phi_{j,m}-\phi_{j,0}$ is the phase shift and
$\hat{J}$ is the Laplacian matrix of the network given by

\begin{equation}
\hat{J}_{ij}=\begin{cases}
-J_{ij}g_{ij}'(\Delta_{ij,0}^{*}) & \mbox{for }i\neq j\\
{\displaystyle \sum_{k,k\neq i}}J_{ij}g_{ik}'(\Delta_{ik,0}^{*}) & \mbox{for }i=j.
\end{cases}
\end{equation}
Now, identifying the matrices $X_{i,m}=D_{i,m}$ and $Y_{i,m}=\theta_{i,m}$
we have reduced the problem of identifying network topology using
distributed perturbations in systems of limit cycle oscillators to
solving the same linear algebraic equation (\ref{eq:linearRestrictionsCORE}). 

\begin{figure}
\centering{}\includegraphics[scale=0.18]{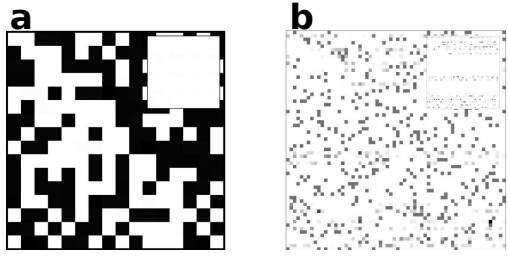}\caption{\textbf{Revealing network topologies from response dynamics for}\emph{
}\textbf{\textcolor{black}{directed networks of phase-locked Kuramoto
oscillators.}}\textcolor{black}{{} Variables evolve according to $\dot{x}_{i}=\omega_{i}+k^{-1}\sum_{j=1}^{N}J_{ij}\sin(x_{j}-x_{i})$,
with random frequencies $\omega_{i}\in[0.1,1]$ and $k=8$ directed
interactions of strengths $J_{ij}=k^{-1}$,} randomly selected for
each unit. Panels show reconstructed coupling matrices $J$\emph{
}for (a) $N=16,$ and $M=32$, and (b) $N=64,$ and $M=32$\textbf{\emph{.}}
The matrices are gray-coded from white ($\hat{J_{ij}}$=0) to black
($\hat{J}_{ij}=\max_{i'j'}\{\hat{J}_{i'j'}\}$). Insets: Element-wise
absolute difference $|J_{ij}^{\text{derived}}-J_{ij}^{\text{original}}|$,
plotted on the same scale.\label{fig:Revealing-network-topologiesKuramotoMatrices}}
\end{figure}

As remarked in previous sections, several experiments are necessary
in order to perform the reconstruction of $\hat{J}$. Therefore, from
repeated measurements for different conditions it is possible to rewrite
eq. (\ref{eq:DistPert3}) in the form (\ref{eq:linearRestrictionsCORE})
in terms of $Y=D\in\mathbb{R}^{N\times M}$ and $X=\Theta\in\mathbb{R}^{N\times M}$
representing the differences between collective dynamics and phase
shifts for each of the $N$ systems during the $M$ experiments. 

\begin{figure}
\centering{}\textbf{\includegraphics[scale=0.14]{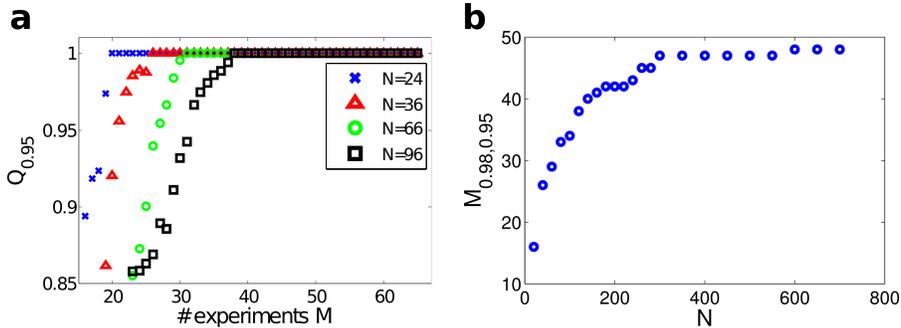}}\caption{\textbf{Quality of reconstruction and number of required experiments
for reconstructing}\emph{ }\textbf{\textcolor{black}{directed networks.}}\textcolor{black}{{}
Phase-locked Kuramoto oscillators with dynamics defined as in Fig.~\ref{fig:Revealing-network-topologiesKuramotoMatrices}
with} $k_{i}=10$ random incoming connections per node\textbf{\emph{.}}\emph{
}(a) Quality of reconstruction at $\alpha=0.95$ for $N=24$ ($\times$),
$N=36$ ($\vartriangle$), $N=66$ ($\circ$) and $N=96$ ($\square$).
(b) Minimum number of experiments required for a reconstruction of
quality level $q=0.98$ and accuracy $\alpha=0.95$.\textbf{ \label{fig:Quality-of-reconstruction}}}
\end{figure}

Now, for sufficiently many experiments, i.e. $M\geq N$, the reconstruction
may be accomplished in principle but may be ill-conditioned numerically.
In addition, for large $N$ the many required experiments may not
be practical. if this condition is not fulfilled, methods like the
\emph{setting a maximum connectivity per system }or \emph{maximizing
the sparseness of the connectivity matrix }(section \ref{sub:Maximizing-the-sparseness})\emph{,
}may be applied in order to make the problem of retrieving $\hat{J}$
an overdetermined problem.

As shown in \cite{Timme:2007p14319}, we may compare how accurate
our prediction is by defining $J_{\max}:=\max_{i'j'}\left\{ \left|J_{i'j'}^{\text{derived}}\right|,\left|J_{i'j'}^{\text{original}}\right|\right\} $,
and using a relative difference defined as
\begin{equation}
\Delta J_{ij}:=\dfrac{1}{2J_{\max}}\left|J_{ij}^{\text{derived}}-J_{ij}^{\text{original}}\right|,\label{eq:DistPert3.5}
\end{equation}
where $\Delta J_{ij}\in\left[0,1\right]\,\forall\, i,j$ . In addition,
the quality of reconstruction $Q_{\alpha}$ may be posed as the fraction
\begin{equation}
Q_{\alpha}:=\dfrac{1}{N^{2}}\sum_{i,j}H\left((1-\alpha)-\Delta J_{ij}\right)\in\left[0,1\right]\label{eq:DistPert4}
\end{equation}
of connection strengths which are assumed to be correct. Here $\alpha\leq1$
is a constant employed to set the required accuracy for predictions
and $H$ is the Heaviside function, $H(x)=1$ for $x\geq0$ and $H(x)=0$
for $x<0$. For instance, $\alpha=0.95$ means that the derived matrix
has a normalized relative error \eqref{eq:DistPert3.5} of at most
5\%. Moreover, we may estimate the minimum number of experiments 
\begin{equation}
M_{q,\alpha}:=\min\left\{ M|Q_{\alpha}(M)\geq q\right\} \label{eq:DistPert5}
\end{equation}
required for a reconstruction with a quality level $q$ and with a
prediction accuracy $\alpha$. Figure \ref{fig:Quality-of-reconstruction}
illustrates these measures for random networks of phase-locked Kuramoto
oscillators for several random topologies and parameters.

The driving response method in principle may be applied to a broad
variety of problems involving stable dynamics. A model analogous to
eq. (\ref{eq:DistPert3}) could be inferred as long as the systems
may be linearized around a stable state, allowing to retrieve the
topology from the network responses as above. Yet, there may be practical
problems. For instance, even for perturbations induced by constant
driving signals, the invariant solution resulting from perturbations
to more complex periodic orbits or other stable invariant sets may
be describable only by time dependent quantities (and not, e.g. temporally
constant phase differences), limiting the approach suggested above
to specific classes of systems.

\subsection{Features and restrictions}

One common advantage of the approaches presented above is that their
required computational effort scales well (weaker than linearly) with
system size $N$ such that at least moderately large systems appear
accessible (cf. Figure \ref{fig:Quality-of-reconstruction}). At the
same time, the approaches are relatively simple to realize because
they do not require knowledge in higher mathematics or computational
approaches beyond a basic standard. 

A possible route of generalization is to combine some of the above
approaches. For instance, one may first drive a system to a stable
fixed point as in\textbf{ }section \ref{sub:Dynamics-Towards-FixedPoints}
and then apply small perturbations around that new point as in section
\ref{sub:Driving-the-system's}.

Yet, all these approaches require the researchers to be able to access
(measure and drive) the dynamics of all units in the system. Moreover,
the local dynamics as well as the (approximate) form of interactions
typically need to be at least partially known. The collective dynamics
suitable for the driving-response approaches described above also
need to be simple, in fact to exhibit a stable fixed point or periodic
orbit or to admit the system to be driven to such as state. Finally,
the presented inference of the existence of physical interactions
and their functional form \cite{Zoran:2011} seems well understood
for networks of phase-oscillators, where perturbations in oscillation
amplitude decays on faster time scale than the relaxation of phases.
It thus remains an open problem how to use a driving response approach
to properly infer structural network connectivity of coupled oscillators
in systems, where the amplitude degrees of freedom play a role or
are even dominant. More generally, systems exhibiting more complex
dynamics, such asynchronous chaotic activity, bifurcations, multistability
or other prevalent features of high-dimensional, nonlinear systems,
currently still prevent network reconstruction by the methods presented
above. 

These requirements severely restrict the range of applicability in
praxis to simple, well-accessible systems only. In particular for
biological systems such as neural circuits or gene interaction networks,
dynamics are typically more complex, systems are large and it is still
hard to implement controlled large-scale driving experiments on the
single-unit level. Direct methods (section \ref{sec:Direct-approaches})
that do not rely on driving the system seem to offer viable directions
towards reconstructing networks with such more complex dynamics as
well. A currently open question of research constitutes how to exploit
recorded time series from only a fraction of the units.

\section{Copy-Synchronization: Adapting a model copy \label{sec:Synchronizing-a-Model}}

Another way of reconstructing network topology of a given network
is by adapting the topology of a\emph{ }second, model system, a\emph{
network} \emph{copy,} such that it synchronizes with the original
system. The idea is to update the model topology continuously until
the copy system exhibits a dynamics identical to the original system;
the rationale is that the final topology of the copy is likely to
be the original topology \cite{Yu:2006p13046}.

\begin{figure}[h]
\begin{centering}
\includegraphics[scale=0.061]{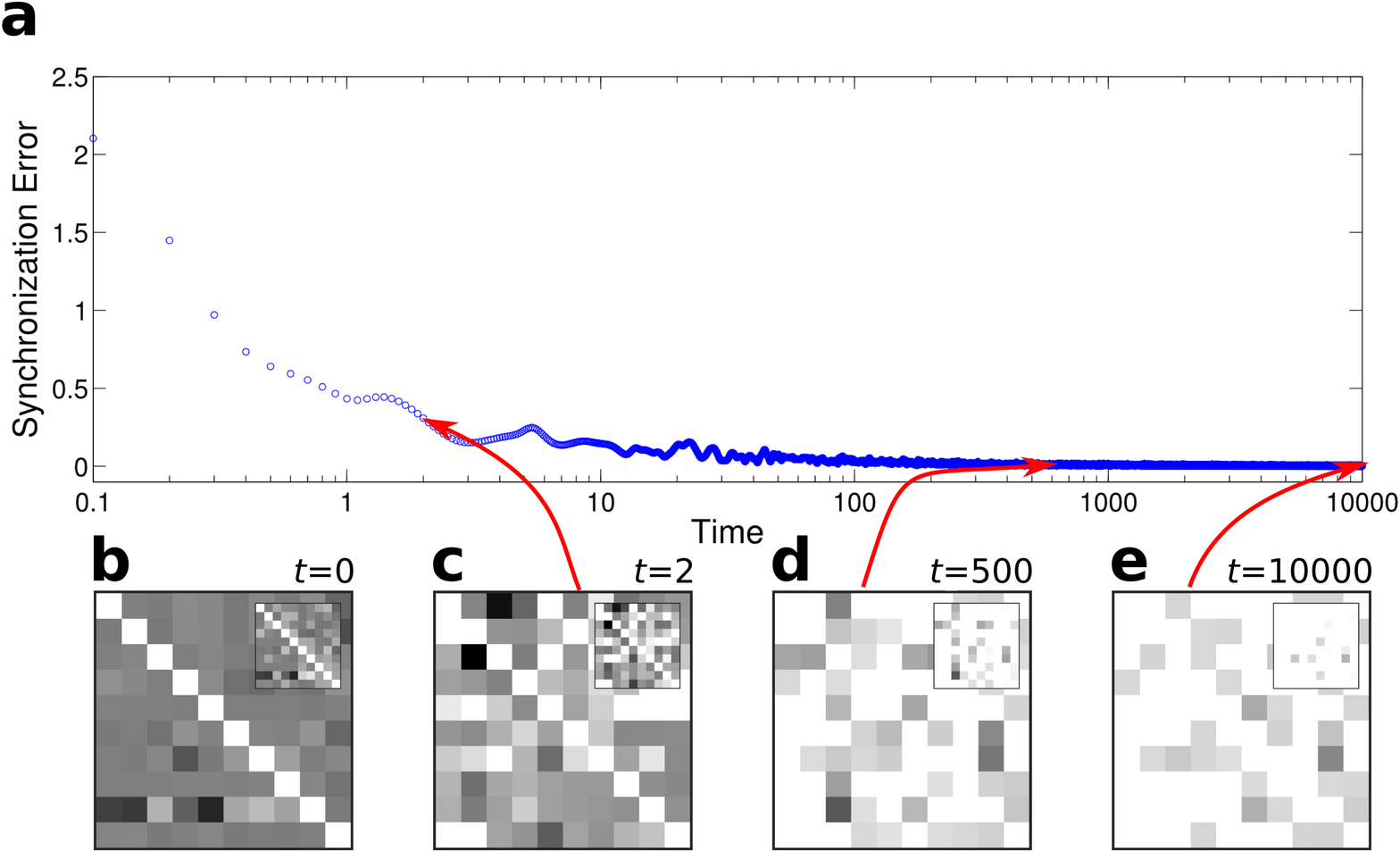} 
\par\end{centering}

\caption{\textbf{\textcolor{black}{Revealing network topology via Copy Synchronization.
}}\textcolor{black}{Dynamics of reconstructed network for directionally
coupled Kuramoto oscillators $\dot{x}_{i}=\omega_{i}+k^{-1}\sum_{j=1}^{N}J_{ij}\sin(x_{j}-x_{i})$
with random frequencies $\omega_{i}\in[-1,1]$, random coupling strengths
$J_{ij}\in[0.5,1]$, $N=10$ and $k=3$ random incoming connections
per node}\textbf{\textcolor{black}{\emph{.}}}\textcolor{black}{\emph{
}}\textcolor{black}{(a) Dynamics of synchronization error between
the original and the copy network measured as $e(t)=\sqrt{N^{-1}\sum_{i}e_{i}^{2}(t)}$.
(b), (c), (d) and (e) show the inferred topologies at $t=0$, $t=2$,
$t=500$ and $t=10^{4}$. }The matrices are gray-coded from white
($J_{ij}$=0) to black ($J_{ij}=1$). The insets depict the element-wise
absolute differences $|J_{ij}^{\text{derived}}-J_{ij}^{\text{original}}|$,
plotted on the same scale.\textcolor{black}{{} }\textcolor{green}{\label{fig:Revealing-network-topology}}}
\end{figure}

Specifically, consider an (original) system of the form
\begin{equation}
\dot{x}_{i}=f_{i}(x_{i})+\sum_{j=1}^{N}J_{ij}g_{j}(x_{j}),\label{eq:SynchMod1}
\end{equation}
where $f_{i}$ and $g_{ij}$ are known and assumed to be Lipschitz
continuous. This original system can in principle have arbitrary dynamics.
Now, let us propose a model copy

\begin{equation}
\dot{y}_{i}=f_{i}(y_{i})+\sum_{j=1}^{N}K_{ij}g_{j}(y_{j})+I_{i},
\end{equation}
where $y_{i}$ represents the state of the copy system, $I_{i}(t)$
are control feedback signals and $K_{ij}(t)$ is current coupling
strength in the test system. In order to synchronize the copy to the
original system, both the feedback signals $I_{i}(t)$ and the coupling
strengths $K_{ij}(t)$ evolve in time and depend on the states of
the actual and the copy system. Defining the synchronization error
\begin{equation}
e_{i}=y_{i}-x_{i}\label{eq:SynchronizationError}
\end{equation}
we adapt the coupling strengths in the model copy according to 

\begin{equation}
\dot{K}_{ij}=-\gamma_{ij}g_{j}(y_{j})e_{i}\label{eq:SynchMod2}
\end{equation}
and

\begin{equation}
I_{i}=-\alpha e_{i}\label{eq:SnychMod3}
\end{equation}
where $\alpha>0$ and $\gamma_{ij}>0$. We remark that here the local
dynamics $f_{i}(.)$ as well as the coupling function $g_{ij}(.)$
need to be known in order to set up the test copy. It was proven in\emph{
}\cite{Yu:2006p13046} that under feedback signals (\ref{eq:SnychMod3})
with sufficiently large $\alpha$, the synchronization error $e_{i}$
decreases in time, $\dot{e}_{i}\leq0$ for all $i$ such that the
two systems converge to a synchronized state. The rationale is that
after synchronization, the copy topology equals that of the original
network, $K_{ij}\approx J_{ij}$, cf. Fig. \ref{fig:Revealing-network-topology}.
With minor modifications on the control signals $I_{i}$ this method
admits to reconstruct networks and sub-networks in the presence of
disturbances and modeling errors as well \cite{Yu:2006p13046}. 

We remark that to the best of our knowledge, there is no guarantee
that the resulting topology of the copy system actually reflects the
one of the original network. In particular, symmetries might lead
to disambiguities.

Further, the method based on copy synchronization is model dependent
such that knowing the intrinsic and coupling functions of units is
vital, as in several parts of section (\ref{sec:Measuring-the-Response}).
Its applicability has been explicated for sample networks of up to
$N=16$ nodes, but it remains unclear how to handle large networks
as of the order of $N^{2}$ links $K_{ij}$ need to be co-evolved
in time and a bound of convergence times is currently missing.\textbf{
}At the same time, the copy approach does not require perturbations
to the original systems, so experimental access to it need not include
driving access to its units. Interestingly, Yu \emph{et.al }\cite{Yu:2006p13046}
highlight that the method may be useful to track changes in a network's
connectivity in real time.

\section{Direct approaches\label{sec:Direct-approaches}}

In the previous two sections, we have introduced methods to infer
network connectivity by either interfering with the system (driving
response approaches, Section \ref{sec:Measuring-the-Response}) or
by setting up and synchronizing a second, model system (section \ref{sec:Synchronizing-a-Model}).
Both classes of methods work if certain requirements are met (in particular,
the option to actively drive the system or the option to synchronize
the copy with the original system, respectively).  It remains a challenge
to infer network topology from dynamics without such requirements.

\subsection{Reconstruction by purely observing dynamics\label{sub:Reconstruction-by-observation}%
\footnote{part of the material presented in this subsection is taken from \cite{Srinivas:2011}and
partially modified; this is not to be confused with the Guttenplag
method.%
}}

Methods based on copy-synchronization (section \ref{sec:Synchronizing-a-Model})
assume that the local dynamics $f_{i}$ as well as the interaction
functions $g_{ij}$ in (\ref{eq:basicHighDimNetwork}) are known and
the $g_{ij}$ do not depend on the state of unit $i$, $g_{ij}(x_{j})$.
Inferring network connectivity, i.e. the $J_{ij}$, then relies on
the construction of a second, model network, with dynamics governed
by Eq.~\ref{eq:basicHighDimNetwork} and network parameters $J'_{ij}$
that are tuned to that of the real network by an error minimization
procedure. As noted recently \cite{Srinivas:2011}, one may solve
the same reconstruction problem with significantly reduced efforts
and reduced requirements by evaluating the states and their time derivatives
directly from the time series recorded from the original system. In
particular, such a simple direct method \cite{Srinivas:2011} removes
the need to set up and synchronize a second, copied, system.

The idea is as follows: If the local dynamics and the coupling functions
are known, their evaluations at different times are also known from
recorded time series and the only remaining unknown parameters in
Eq.~\ref{eq:basicHighDimNetwork} are the coupling strengths, which
are to be determined. Specifically recording a time series $x_{i}\left(t_{m}\right)$
at sufficiently closely spaced times $t_{m}\in\mathbb{R}$ and estimating
the temporal derivatives%
\footnote{We remark that equidistant times $t_{m}$ of sampling are not necessarily
the best to evaluate $\dot{x}_{i}$, in particular if the time derivatives
are approximated linearly.%
} of it yields the dynamics of the \emph{i}-th unit given by

\begin{equation}
\dot{x}_{i}\left(t_{m}\right)=\mathit{f_{i}\left(x_{i}\left(t_{m}\right)\right)+\sum_{j=1}^{N}J_{ij}g_{ij}\left(\mathit{x{}_{i}\left(t_{m}\right),x}_{j}\left(t_{m}\right)\right).}\label{eq:general-equation-2}
\end{equation}
If there are \emph{$M$} such times, $m\in\left\{ 1,\ldots,M\right\} $,
we have \emph{M} equations of the form

\begin{equation}
\dot{x}_{i,m}=f_{i,m}+\sum_{j=1}^{N}J_{_{ij}}g_{ij,m}\,\label{eq:DEshortform}
\end{equation}
with abbreviations $\dot{x}_{i,m}:=x_{i}(t_{m})$, $f_{i,m}:=f_{i}\left(x_{i}\left(t_{m}\right)\right)$
and $g_{ij,m}:=g_{ij}\left(\mathit{x{}_{i}\left(t_{m}\right),x}_{j}\left(t_{m}\right)\right)$.
Repeated evaluations of the equations of motion (\ref{eq:general-equation-2})
of the system at different times $t_{m}$ thus comprise a simple and
implicit restriction on the network topology $J_{ij}$ as follows:
writing $Y_{i,m}=\dot{x}_{i,m}-f_{i,m}$ and the matrix $X_{i}=(g_{ij,m})_{j,m}\in\mathbb{R}^{N\times M}$,
these equations constitute the matrix equation 
\begin{equation}
\boldsymbol{Y}_{i}=\boldsymbol{J}_{i}X_{i}\label{eq:DirectComponentmatrix}
\end{equation}
where $\boldsymbol{Y}_{i}\in\mathbb{R}^{1\times M}$ and $\boldsymbol{J}_{i}\in\mathbb{R}^{1\times N}$
is the \emph{i}-th row of the interaction matrix $J$, comprising
the vector $(J_{ij})_{j\in\{1,\ldots,N\}}$ of all input coupling
strengths to unit \emph{i}. 

\begin{figure}[h]
\centering{}\includegraphics[scale=1.5]{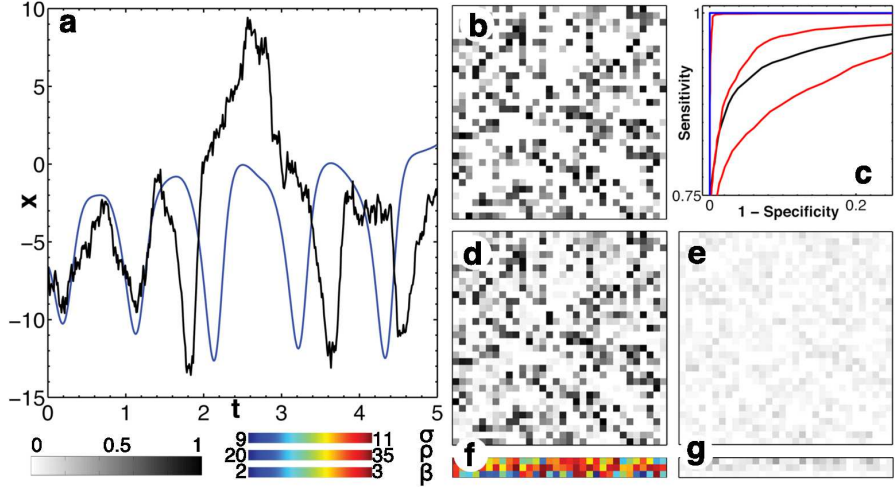}\caption{\textbf{Reconstructing a random network complex chaotic dynamics.
}Example of $N=32$ Lorenz oscillators with dynamics given by $\dot{x}_{i}=\sigma_{i}(y_{i}-x_{i})+\sum_{j=1}^{N}J_{ij}(x_{j}-x_{i})+\xi_{i}^{(x)},\,\dot{y}_{i}=x_{i}(\rho_{i}-z_{i})-y_{i}+\xi_{i}^{(y)},\,\dot{z}_{i}=x_{i}y_{i}-\beta_{i}z_{i}+\xi_{i}^{(z)}$,
with unknown parameters $\forall\, i,j\,\,\{J_{ij},\sigma_{i},\rho_{i},\beta_{i}\}$
in the presence of external Gaussian white noise $\boldsymbol{\xi}_{i}$
of amplitude $\lambda$\textbf{\emph{.}}\emph{ }(a) Dynamics of a
unit in the absence (blue) and presence (black) of noise with $\lambda=5$.
Original (b) and reconstructed (c) connectivity, inferred parameters
(f), and their errors compared to the original topology and parameters
(e) and (g). (c) Receiver operating characteristics (ROC) of reconstruction
under noise-free (blue) and noisy measurements (black and red) for
$\lambda\in\{0.1,1,10\}$. Figure taken from \cite{Srinivas:2011}.
\label{fig:ReconstructionNoise}}
\end{figure}

The main restricting equations (\ref{eq:DirectComponentmatrix}) again
have the same form as the generic restrictions (\ref{eq:linearRestrictionsCORE})
and thus may be solved analogously. In \cite{Srinivas:2011},\textbf{
}Euclidean $L_{2}$-norm minimization was used to infer the topology.\textbf{
}Numerical tests show that reconstruction works well for transient
as well as attractor dynamics, for simple as well as complex, chaotic
units and collective states, and even in the presence of noise that
substantially alters the dynamics and thus the recorded time series.
Figure \ref{fig:ReconstructionNoise} illustrates successful reconstruction
in the presence of various levels of noise. 

It is clear that certain types of dynamics do not allow topology inference.
For instance, if all local dynamics are identical $f_{i}\equiv f$,
all coupling functions are identical, $g_{ij}\equiv g$, and collectively
the units are identically synchronous, i.e. all in the same states
at the same times, there is no information about the connection topology
one can possibly obtain from (only) recording time series, because
for \emph{all} strongly connected topologies%
\footnote{A network is strongly connected if for every pair of nodes $(i,j)$,
node $j$ can be reached from node $i$ via a (directed) path within
the network \cite{Timme:2008,Diestel:2010}.%
}, the collective dynamics would be identical. 

Further theoretical considerations show that following the same lines
of analysis as above, all parameters occurring linearly in the local
dynamics of coupling functions, can also be reconstructed by the same
error minimization based on eqn.~(\ref{eq:linearRestrictionsCORE}).
For instance for the (fictional) coupling function $g_{ij}(x,y)=a\sin(xy)-\exp(b+y-x)$,
the parameters $a$ and $b$ need not be known but can be inferred
as well (up to one constant prefactor for each pair $(i,j)$ of nodes),
because both $a$ and $\exp(b)$ occur linearly in the equations of
motions (\ref{eq:general-equation-2}). In many physical systems,
parameters actually do occur linearly. These include, for instance,
the dynamics of coupled classical electric LCR circuits and the strengths
of diffusive (\ref{eq:diffusiveCoupling}) or direct linear coupling
(\ref{eq:directLinearCoupling}). Moreover, widely used model systems
for networks of neurons, such as leaky integrate-and-fire and quadratic
integrate-and-fire neurons \cite{Izhikevich:2007}\textbf{ }or networks
of coupled chaotic systems such as R\"{o}ssler or Lorenz systems
\cite{Rossler:1976p15735,Lorenz:1963p15795} exhibit all or at least
most parameters occurring linearly or affinely. For all such systems,
local dynamics $f_{i}$ and coupling functions $g_{ij}$ may not or
not completely be required to be known in advance.

Such a direct inference method \cite{Srinivas:2011} for network topology
from complex collective dynamics thus serves as a simple starting
strategy for connectivity reconstruction in a broad range of networked
systems.

\subsection{Pulse-coupling: Reconstruction from spike patterns\label{sub:Reconstruction-from-spike}}

Networks of spiking neurons or other pulse-coupled units are formally
hybrid systems \cite{Aswhin:2005}, because the continuous-time flow
is interrupted at certain time points, where events (e.g. spike sending
or reception) modify the dynamics \cite{Cessac:2008,Kielblock:2011}\textbf{
}via discrete time maps. 

To start addressing the reconstruction problem for such hybrid systems,
researchers have considered one of the simplest model networks of
pulse-coupled units based on leaky integrate-and-fire (LIF) models.
Such network models are most broadly used in mathematical analysis
and computational modeling. Each unit in such a network has one state
variable, its membrane potential, roughly emulates leaky capacitive
properties observed for membranes of biological neurons and is complemented
with a threshold where a pulse (action potential or spike \cite{Rieke:1997})
is artificially created and the membrane potential is reset. Although
these models lack certain biological details, such as a natural spike
generating mechanisms, they are simple enough to be studied analytically
and they have been useful for furthering our understanding how the
information is processed among neurons \cite{Burkitt:2006}. 

Explicitly, the membrane potential $V_{i}(t)$ of a LIF unit $i$
changes as in an RC circuit (resistor and a capacitor connected in
parallel) according to 
\begin{equation}
\dot{V}_{i}=\gamma_{i}(R_{i}I_{i}-V_{i})+S_{i}(t),\label{eq:bussel1}
\end{equation}
where $R_{i}$, $\gamma_{i}$ and $I_{i}$ are the membrane resistance,
inverse membrane time constant and the external current applied to
neuron $i$, respectively. Once the potential of a unit $j$ crosses
a threshold, $V_{j}(t)\geq V_{T,j}$, the potential is reset to $V_{j}(t^{+})=V_{R,j}$
and the unit emits a pulse that it transmitted to the neuron's post-synaptic
neighbors \cite{TheoreticalNeuro2001}. The time of this pulse sending
event is remembered as the unit's $m$-th spike time $t_{j,m}:=t$.
The collection of such pulses then define the actions onto post-synaptic
units $i$ such that the quantity 
\begin{equation}
S_{i}(t)=\sum_{j=1}^{N}\sum_{k\in\mathbb{Z}}J_{ij}\delta(t-t_{j,m}-\tau_{ij}),
\end{equation}
in \eqref{eq:bussel1} represents the pulses that unit $i$ receives
from the rest of the network. Here, $J_{ij}$ and $\tau_{ij}$ are
the synaptic coupling strength and the synaptic transmission delay
from unit $j$ and $i$, respectively. Furthermore, $\delta(.)$ is
the Dirac delta distribution modeling a potential response kernel
that is much faster than all other time scales involved.

The main question we address now is whether and how the network connectivity,
as specified by the matrix of coupling strength $J_{ij}$ can be reconstructed
if the pattern of spikes times $\left(t_{j,m}\right)_{j\in\{1,\ldots,N\},m\in\mathbb{Z}}$
is given? We remark that we do not assume access to the natural state
variables, the membrane potentials $V_{i}(t)$, which may not be observable,
but only to the spike times that are generated by the dynamics of
these potentials. This difference constitutes the main novelty of
the approaches presented in this subsection compared to those for
continuous time state variables presented before.

First observe that (\ref{eq:bussel1}) has an explicit solution \cite{VanBussel:2011}
given by 

\begin{equation}
V_{i}(t)=R_{i}I_{i}(1-e^{-\gamma_{i}(t-t_{0})})+V_{i}(t_{0})e^{-\gamma_{i}(t-t_{0})}+\sum_{j=1}^{N}\sum_{t_{0}<t_{j,m}+\tau_{ij}\leq t}J_{ij}e^{-\gamma_{i}(t-t_{j,m}-\tau_{ij})}.\label{eq:bussel2}
\end{equation}
if the time interval $[t_{0},t)$ lies in between two subsequent spikes
of neuron $i$. The first sum is taken over all neurons while the
second is taken over those spikes generated by the other neurons $j\neq i$
that affect $i$ in the time interval $[t_{0},t)$.

Based on this solution, we now present two distinct approaches to
infer network connectivity, one direct and exact inference method
assuming oscillatory units and one coarse approximate method based
on stochastic optimization:

\subsubsection{Direct topology inference assuming oscillatory units}

Van Bussel et al. \cite{VanBussel:2011} assumes that all the parameters
of (\ref{eq:bussel2}), besides the synaptic couplings $J_{ij}$,
are known. The rationale is that delays and membrane time constants
as well as a unit's equilibrium potential $R_{i}I_{i}$ may be estimated
beforehand. How could these coupling strengths and thus the structural
network topology be inferred? If the currents $I_{i}$ are sufficiently
large such that $R_{i}I_{i}>V_{T,i}$ the units are oscillatory such
that they create spike even without recurrent inputs from the remaining
network. After crossing the threshold $V_{T,i}$, unit $i$ sends
a spike to its post-synaptic neighbors and is reset to a resting value
$V_{R,i}$, so that the unit's state at the boundaries of each inter-spike
interval $[t_{i,k-1},t_{i,k})$ is determined and one may take $t_{0}=t_{i,k-1}$
and $t=t_{i,k}$. However, these threshold crossings may be induced
in two manners at $t=t_{i,k-1}$, (a) by incoming excitatory spikes
from other neurons such that $V_{i}(t^{-})<V_{T,i}$ but $V_{i}(t^{-})+\sum_{\left\{ j:\exists m:t_{j,m}+\tau_{ij}=t\right\} }J_{ij}>V_{T,i}$
or (b) by the intrinsic oscillation\emph{ }of the unit such that $V_{i}(t^{-})=V_{T,i}$
without simultaneously incoming spike(s). In both cases, the membrane
potential is at reset value immediately after sending a spike. So
identifying $t_{0}=t_{i,k-1}$ in (\ref{eq:bussel2}), we have 
\begin{equation}
\begin{array}{c}
V_{i}(t_{0}):=V_{i}(t_{i,k-1})=V_{R,i}.\end{array}\label{eq:bussel3}
\end{equation}
If the next spike is oscillation-induced (b), the membrane potential
approaches its threshold value $V_{T,i}$ continuously such that in
addition

\begin{equation}
\begin{array}{c}
V_{i}(t^{-}):=\begin{array}{c}
V_{i}(t_{i,k}^{-})=V_{T,i}\end{array}\end{array}\label{DUPLICATE: eq:bussel4}
\end{equation}
where we now identified $t=t_{i,k}$ in (\ref{eq:bussel2}). Thus,
each oscillation-induced spike at some $t=t_{i,k}$ implies a linear
restriction of the form (\ref{eq:bussel2}) for the coupling strengths
$J_{ij}$ by equating $[t_{0},t)=[t_{i,k-1},t_{i,k})$. 

For $M$ different inter-spike intervals obeying (\ref{eq:bussel3})
and (\ref{DUPLICATE: eq:bussel4}), this provides a linear system
of equations restricting the coupling matrix. However, as van Bussel
et al. \cite{VanBussel:2011} remark, consecutive intervals may display
similar patterns, such that it is often advisable to select $M>N$
inter-spike intervals sufficiently separated in time, to minimize
correlations between intervals and thus numerical inaccuracies. For
each $i$, defining the subset of inter-spike intervals 
\begin{equation}
D_{i,m}:=(t_{i,l_{m}-1},t_{i,l_{m}})\label{eq:ChosenIntervals}
\end{equation}
 and 
\begin{equation}
T_{i,m}=t_{i,l_{m}}-t_{i,l_{m}-1}=|D_{i,m}|\label{eq:ChosenIntervalDuration}
\end{equation}
eq.~(\ref{eq:bussel2}) may be rewritten as 

\begin{equation}
X_{i}\boldsymbol{J}_{i}^{T}=\boldsymbol{Y}_{i},
\end{equation}
where $\boldsymbol{J}_{i}:=(J_{ij})_{j\in\{1,\ldots,N\}}$ and $X_{i}\in\mathbb{R}^{M\times N}$
and $\boldsymbol{Y}_{i}\in\mathbb{R}^{M\times1}$ are given by

\begin{equation}
\left(X_{i}\right)_{mj}=\sum_{t_{j,k}+\tau_{ij}\in D_{i,m}}J_{ij}e^{-\gamma_{i}(t_{i,l_{m}}-t_{j,k}-\tau_{ij})}
\end{equation}
 and

\begin{equation}
Y_{i,m}=V_{T,i}-R_{i}I_{i,m}(1-e^{-\gamma_{i}T_{i,m}})-V_{R,i}e^{-\gamma_{i}T_{i,m}}.
\end{equation}

Repeating the same process for all units $i$ yields the topology
of the whole network. As shown in section \ref{sub:Solving-the-restricting-equations}
, the overdetermined problem, $M>N$, and the undetermined, $M\ll N$,
may be solved minimizing the $L_{2}$-norm or maximizing the sparseness
of the network, respectively. A major limitation is that the presented
approach requires a large amount of prior knowledge about the system,
such as the time delays between units and their time constants, among
others. Given this knowledge, the approach is capable of inferring
large networks of neurons as the computation time, as well as the
number $M$ of required inter-spike intervals to be evaluated scales
linearly with system size\textbf{ }\cite{VanBussel:2011,vanBussel:2010}.

\subsubsection{Stochastic optimization of all systems parameters}

In a complementary study, Makarov et al.~\cite{Makarov:2005p9589}
showed how stochastic optimization of all system parameters 
\begin{equation}
\boldsymbol{p}_{i}=\left(\gamma_{i},R_{i},I_{i},\tau_{ij},J_{ij}\right)_{i,j\in\{1,\ldots,N\}}\label{eq:allParameters}
\end{equation}
given the spike trains for all neurons in the network may yield a
network topology roughly consistent with the actual one.

The idea is to evaluate the predicted inter-spike interval durations
$\bar{T}_{i,m}(\boldsymbol{p})$ from a test set of parameters and
to optimize those parameters to reproduce a given (measured) spike
train as closely as possible. Thus, minimizing the difference between
the predicted and measured spike trains is vital for this method.
The authors in particular applied their idea also to recordings of
biological neurons.

As the eq. (\ref{eq:bussel2}) yields transcendental relations for
the $\bar{T}_{i,m}(\boldsymbol{p})$, their estimates need to be determined
numerically. Thus, finding

\begin{equation}
\boldsymbol{p}_{i}^{*}=\arg\min_{\boldsymbol{p}_{i}}\, E_{i}(\boldsymbol{p}_{i})=\arg\min_{\boldsymbol{p}_{i}}\sum_{m=1}^{M}(T_{i,m}-\bar{T}_{i,m}(\boldsymbol{p}_{i}))^{2}\label{eq:stochasticMIN}
\end{equation}
yields the best (in Euclidean distance norm) solution for the set
of parameters.

Briefly, to find the minimum $\boldsymbol{p}_{i}^{*}$ in (\ref{eq:stochasticMIN})
one must explore the parameters space. This means that the solution
is found through iteratively choosing random values for $\boldsymbol{p}_{i}$
and comparing the value of (\ref{eq:stochasticMIN}) for consecutive
iterations. By relating the changes in (\ref{eq:stochasticMIN}) with
the changes in $\boldsymbol{p}_{i}$ one may establish directions
in which the minimum may be found by gradient descent. However, it
is also advisable to change directions in the parameters space randomly.
Mainly, because there may better solutions for $\boldsymbol{p}_{i}^{*}$
in regions of the parameters space where they are not expected to
exist \cite{Gentle:2004}. Makarov et al.~\cite{Makarov:2005p9589}
thus applied stochastic optimization for searching the minimum.

As a strong requirement, this method needs that the number of recorded
spikes to be $M$$\gg2N$; as noted in \cite{Makarov:2005p9589},
robust regression models are more suitable to handle this kind of
problems and special care with the inter-spike intervals must be taken
as, e.g., spike bursts may lead to false intervals.

\subsection{Features and limitations}

The stochastic optimization approach provides a generic approach in
finding best fitting parameters and thus here, potentially a well
fitting network; yet, it is computationally demanding and simultaneously
requires many recordings compared to the size of the network. In contrast,
the direct approach assumes a large amount of pre-knowledge, in particular
regarding the unit's parameters. In addition, by requiring to pre-process
the data sets (i.e. choosing appropriate time intervals), the minimization
problem is no longer required to deal with transcendental relations.
This leads to a considerable increase on the computational performance
and it is the key factor that makes the method suitable for large
networks. We remark that still, both methods are currently not suitable
in our opinion to reliably infer network topology from real recordings
of spike data, because of several reasons: For instance, other network-external
sources of spike generation, stochastic fluctuations due to intrinsic
noise and the highly specific conditions required for reconstruction
still limit the methods applicability. Finally, both approaches assume
linearity of interactions, yet it is known that interactions can be
nonlinear, e.g. due to dendritic spikes in response to sufficiently
strong, simultaneous inputs to single dendritic branches of a neuron
\cite{Schiller:2003,Memmesheimer:2012,Jahnke:2012}.

\section{Technical issues\label{sec:Technical-issues}}

Let us briefly comment on four technical issues related to the structural
inference approaches discussed above. We mention how some of them
may be directly transferred to settings, where discrete time dynamics
describes the system (section \ref{sub:Discrete-time-maps}), discuss
issues when measuring data from real, e.g. biological networks (section
\ref{sub:SamplingRealNetworks}), remark on $L_{1}$ vs. $L_{2}$
minimization (section \ref{sub:L1L2normMinimization}) and finally
discuss what happens for hyper-networks, where more than two-point
interactions influence the collective state (section \ref{sub:Hypernetworks:-beyond-two-point}).

\subsection{Discrete time maps\label{sub:Discrete-time-maps}}

If the dynamical system is described as a network of coupled discrete-time
maps instead of by coupled differential equations (\ref{eq:basicLowDimNetwork}),
approaches similar to the ones presented above are often viable in
slightly modified form. For instance, the core equation becomes (for
1-dimensional local units)

\begin{equation}
\boxed{x_{i}(t+1)=f_{i}\left(x_{i}(t)\right)+\sum_{j=1}^{N}J_{ij}g_{ij}\left(x_{i}(t),x_{j}(t)\right)+I_{i}(t)+\xi_{i}(t)}\label{eq:basicLowDimNetworkMAPS}
\end{equation}
where now the time $t\in\mathbb{Z}$ is an integer. Here, the direct
method of subsection \ref{sub:Reconstruction-by-observation} is immediately
applicable, even with the additional advantage that no temporal derivatives
need to be estimated such that an observed time series $(x_{i}(t))_{i\in\{1,\ldots,N\},t\in\{1,\ldots,M\}}$
enters the inference equations without approximations. Of course,
that time series may still contain substantial measurement errors
that propagate into any solution of the inference problem.

\subsection{Sampling dynamics of real networks\label{sub:SamplingRealNetworks}}

When sampling real life phenomena, measurement constraints may arise
due to the nature of the given phenomena, making the study of such
a major challenge. For instance, in \emph{EEG analysis}, a group of
sensors is employed to measure the brain activity. These recordings
help in finding functional connections among regions in the brain.
Basically, each sensor measures the activity of a population of neurons
below its active area for a certain amount of time. Then, by analyzing
the correlations among the time series of each sensor, the interaction
network is constructed. Nevertheless, as remarked by Bialonski in
\cite{Bialonski:2012}, special considerations regarding the \emph{spatial
}and \emph{temporal sampling} must be accounted for. A simple constraint
is that sensors placed too close to each other, may record overlapping
activity from neighboring populations, thus increasing the difficulty
to discern between \emph{direct} and \emph{indirect }interactions.
Moreover, if the size and frequency of the sampling fails to capture
the intrinsic time scales, false functional connections in network
may be constructed. Clearly this example is particularly relevant
for inferring effective connectivities, as discussed in section \ref{sec:Correlation-based-methods}.
As an example regarding the time domain, gene and protein interaction
networks often still have a very limited number of sampling points
available \cite{Bansal:2007,Hecker:2009} such that a network inference
problem may become drastically under-determined. Additional information
such as the known existence of certain interactions in such a network
(but not the existence of others) sometimes is available but how precisely
to use this is currently unknown and requires further method development.
So briefly, when studying real networks, special technical considerations
should be accounted for to actually use what it is measured (the data)
for what one wants to know about the system, cf. section \ref{sub:Aims-and-options}.

\subsection{$L_{1}$ norm minimization as a linear program}

Given the optimization problem 
\begin{equation}
\min_{\boldsymbol{y}}\left\Vert A\boldsymbol{y}-\boldsymbol{b}\right\Vert _{1},\label{eq:l1LP1}
\end{equation}
where $A\in\mathbb{R}^{M\times N}$, $\boldsymbol{y}\in\mathbb{R}^{N}$
and $\boldsymbol{b}\in\mathbb{R}^{M}$, (\ref{eq:l1LP1}) may be posed
as \cite{Boyd:2009}

\begin{equation}
\min_{\boldsymbol{s}}\boldsymbol{1}^{\mathsf{T}}\boldsymbol{s}\quad\text{s.t. }\begin{array}{c}
A\boldsymbol{y}-\boldsymbol{b}\preceq\boldsymbol{s}\\
A\boldsymbol{y}-\boldsymbol{b}\succeq-\boldsymbol{s}
\end{array},\label{eq:l1LP2}
\end{equation}
where $\boldsymbol{1}\in\mathbb{R}^{M}$ is a vector of ones and $\preceq$
and $\succeq$ denote entry-wise comparison and $\boldsymbol{s}$
is an auxiliary variable. To solve (\ref{eq:l1LP2}), one has to solve
the linear program
\begin{equation}
\min_{x}\tilde{\boldsymbol{c}}^{\mathsf{T}}\boldsymbol{x}\text{\quad\text{s.t.}}\quad\tilde{A}\boldsymbol{x}=\tilde{\boldsymbol{b}},\label{eq:l1LP3}
\end{equation}
where

\begin{equation}
\boldsymbol{x}=\left[\begin{array}{c}
\boldsymbol{y}\\
\boldsymbol{s}
\end{array}\right],\quad\tilde{A}=\left[\begin{array}{cc}
A & -I\\
-A & -I
\end{array}\right],\quad\tilde{\boldsymbol{b}}=\left[\begin{array}{c}
\boldsymbol{b}\\
-\boldsymbol{b}
\end{array}\right],\quad\tilde{\boldsymbol{c}}=\left[\begin{array}{c}
\boldsymbol{0}\\
\boldsymbol{1}
\end{array}\right]\label{eq:l1LP4}
\end{equation}
and $I\in\mathbb{R}^{M\times M}$ and $\boldsymbol{0}\in\mathbb{R}^{N}$
are the identity matrix and a vector of zeroes, respectively. The
advantage of posing problem (\ref{eq:l1LP1}) as (\ref{eq:l1LP3})
is that the latter can be easily solved in a standard way by implementing
any solver for linear programs (e.g. the \texttt{\textcolor{black}{linprog}}
function in MATLAB~\cite{matlab:2006}).

\subsection{$L_{2}\,\text{\,\ vs.\,\,\ \ensuremath{L_{1}}}$ norm minimization\label{sub:L1L2normMinimization}}

The need to choose a minimization scheme may turn the reconstruction
problem into a great challenge, because different schemes may yield
different solutions, thus forcing us to discern which scheme is best
suited for our purposes in specific reconstruction problems. As illustrated
in \cite{Gardner:2003p9614,Timme:2007p14319,Makarov:2005p9589,Napoletani:2008p13092,vanBussel:2010,VanBussel:2011,Yeung:2002}
these differences between minimizers may be exploited in certain situations.
For instance, the $L_{2}$ minimizer finds the closest solution in
the $L_{2}$-norm and moreover has an analytic solution. Yet, given
the nature of the minimizers (check \cite{Napoletani:2008p13092}),
an $L_{1}$ minimizer is suited for finding particular sparse solutions
and it is more robust to outliers than $L_{2}$, so it might be seen
as more useful for applications. However, minimizing an $L_{1}$ norm
is computationally more costly compared to the $L_{2}$ and it may
have more than one solution \cite{Boyd:2009}. We note that $\left\Vert \boldsymbol{p}\right\Vert _{2}\leq\left\Vert \boldsymbol{p}\right\Vert _{1}$
for any vector $\boldsymbol{p}\in\mathbb{R}^{N}$

\subsection{Hypernetworks: beyond two-point interactions\label{sub:Hypernetworks:-beyond-two-point}}

We have explicitly excluded networks with more than two-unit interactions
from our general mathematical description (\ref{sec:Generic-Nonlinear-Dynamics})
or dynamical networks with temporally changing connections. Those
may occur, for instance in networks of computers, or other communication
networks of engineering, where inputs from several units that give
input to the same other unit are nonlinearly processed (for instance,
multiplied) to change the latter units' state. Similarly, non-additive
dendritic interactions in neurons \cite{Memmesheimer:2012,Memmesheimer:2010},
where two simultaneously received synaptic inputs in close spatial
proximity initiate so-called dendritic spikes and thereby a nonlinearity
\cite{Schiller:2003}, naturally imply three-point interactions.

We remark that direct three-unit and higher order interactions (i.e.
three-term and higher order products such as $x_{i}^{d}x_{j}^{d'}x_{k}^{d"}$
where $i,j,k$ are mutually different) are omitted in (\ref{eq:basicHighDimNetwork})
because they refer to dynamical systems on hypernetworks, thus going
beyond the scope of this review. Such third order and higher order
terms are not covered by (\ref{eq:basicHighDimNetwork}), firstly,
because the notation stays much simpler without them, but secondly
and more importantly, because there seem to be few major theoretical
results on reconstructing such systems, if any, that may be or become
of practical use. Yet, in social, communication and information networks,
such questions may soon become of interest as 'big data' are pouring
in. 

A brief introduction to the dynamics of complex hypernetworks is given
in \cite{Timme:2014}.

\section{Effective connectivity: Correlation-based methods\label{sec:Correlation-based-methods}}

\subsection{Linear correlation and covariance}

One common way of quantifying effective connectivity is to measure
scalar time series $\boldsymbol{x}(t)=(x_{1}(t),\ldots,\, x_{N}(t))$,
$t\in\{t_{1},\ldots t_{M}\}$ from the $N$ units of a system (spike
rates of neurons, expression levels of genes, ...), compute their
(temporal) averages $\mu_{i}=\left\langle x_{i}(t)\right\rangle _{t}$
and variances $\sigma_{i}^{2}=\left\langle \left(x_{i}(t)-\mu_{i}\right)^{2}\right\rangle _{t}$
and from those compute the covariance matrix 
\begin{equation}
\mbox{Cov}_{ij}:=\frac{\left\langle \left(x_{i}(t)-\mu_{i}\right)\left(x_{j}(t)-\mu_{j}\right)\right\rangle _{t}}{\sigma_{i}\sigma_{j}}.\label{eq:covariancematrix}
\end{equation}
Here the averages are time averages
\begin{equation}
\left\langle y_{i}(t)\right\rangle _{t}:=M^{-1}\sum_{m\in\{1,\ldots,M\}}y_{i}(t_{m}).\label{eq:timeaverage}
\end{equation}
If multiple measurements are available, averages may be taken over
ensembles of experiments rather than (or in addition to) temporal
averaging. The obtained covariance matrix is then often thresholded,
either just choosing a heuristic value or according to some measure
of statistical significance (against an appropriate null hypothesis)
and the resulting non-zero values interpreted either as ``strength''
of effective connectivity (weighted connectivity matrix) or just as
existence (adjacency matrix) of a functional link between units. Sometimes,
the value of a threshold is systematically varied and changes in resulting
connectivity evaluated. Correlation, covariance and other linear algebra
measures are commonly used, often complemented by nonlinear operations
such as thresholding, to analyze, for instance fMRI or other spatio-temporal
data \cite{Erhardt:2011}. We note that correlation measure may be
(mathematically) seen in a number of different ways \cite{Rodgers:1988}.

\subsection{Entropy maximization}

\emph{Entropy} measures the uncertainty associated with a given probability
distribution and constitutes a key quantity in \emph{information theory
}\cite{Cover:2006}\emph{ .} Given the probabilities of a set of events,
the entropy measures how uncertain, on average, the occurrence of
an event is; or in other words, how much information, on average,
one obtains by measuring the occurrence of that event knowing the
probability distribution of events. Reversely, given a collection
of (observed) data points, we can choose probabilities to maximize
the entropy. Such a distribution, known as a \emph{maximum-entropy
probability distribution,} would be the least biased distribution
possible and any other would require further assumptions on the nature
of the problem \cite{Jaynes:1957}. For a network dynamical system,
i.e. systems of coupled dynamical units, we can ask what the effective
interactions are such that the\emph{ }probability distribution that
best describes the data (averages and correlations) has maximum entropy. 

Specifically, the principle of maximum information from Bayesian statistics
postulates the ``most likely'' probability $\rho(\boldsymbol{x})$
of measuring $\boldsymbol{x}$ given the same type of time series
data $\boldsymbol{x}(t)=(x_{1}(t),\ldots,\, x_{N}(t))$ is the one
maximizing the information obtained from measuring one more state
$\boldsymbol{y}$. More precisely, the goal is to find the probability
$\rho(\boldsymbol{x})$ that maximizes Shannon entropy 
\begin{equation}
S=-\sum_{\boldsymbol{x}}\rho(\boldsymbol{x})\ln\rho(\boldsymbol{x})\label{eq:ShannonEntropy}
\end{equation}
under the constraints that the first and second moments are consistent
with those estimated from the data, 
\begin{equation}
\left\langle x_{i}\right\rangle _{\rho}:=\sum_{\boldsymbol{x}}x_{i}\rho(\boldsymbol{x})\overset{!}{=}\left\langle x_{i}(t)\right\rangle _{t}\label{eq:ensembleAverage1}
\end{equation}
 and 
\begin{equation}
\left\langle x_{i}x_{j}\right\rangle _{\rho}:=\sum_{\boldsymbol{x}}x_{i}x_{j}\rho(\boldsymbol{x})\overset{!}{=}\left\langle x_{i}(t)x_{j}(t)\right\rangle _{t}\label{eq:ensembleAverage2}
\end{equation}
By restricting ourselves to these conditions (and no others), we here
focus on pairwise interactions and neglect three-point and higher
order coupling that arise in hypernetworks (cf. section \ref{sub:Hypernetworks:-beyond-two-point}
and \cite{Timme:2014}). This yields (Appendix \ref{sec:l2minimization})
the probability distribution of the form
\[
\rho(\boldsymbol{x})=Z^{-1}\exp\left(\sum_{i=1}^{N}h_{i}x_{i}+\frac{1}{2}\sum_{i=1}^{N}\sum_{j=1}^{N}\hat{J}_{ij}x_{i}x_{j}\right)
\]
where $Z=\int_{\mathbb{R^{N}}}\exp\left(\sum_{i=1}^{N}h{}_{i}x_{i}+\frac{1}{2}\sum_{i=1}^{N}\sum_{j=1}^{N}\hat{J}_{ij}x_{i}x_{j}\right)d^{N}x$
is a normalization constant and $h_{i}$ and $\hat{J_{ij}}$are parameters
to be chosen such that (\ref{eq:ensembleAverage1APP}) and (\ref{eq:ensembleAverage2APP})
hold. The quantities $\hat{J}_{ij}$ are interpreted as effective
couplings between units $i$ and $j$. 

If the matrix of covariances between the $N$ time series is 
\begin{equation}
C_{ij}=\left\langle x_{i}(t)x_{j}(t)\right\rangle _{t}-\left\langle x_{i}(t)\right\rangle _{t}\left\langle x_{j}(t)\right\rangle _{t}\,.\label{eq:Correlationdefined}
\end{equation}
the effective coupling matrix is given by its inverse (Appendix \ref{sec:l2minimization})\textbf{
\begin{equation}
\hat{J}=C^{-1}.\label{eq:MaxEntropyCoupling}
\end{equation}
}This means that the best available probability distribution (i.e.
that yielding the maximum information) is given by second order effective
coupling strengths determined by (but by no means identical to) the
linear correlation matrix.

This concept is applied to a range of different systems, in particular
in biology, including gene networks \cite{Braustein:2008}, protein
networks \cite{Tkacik:2006} and neural circuits \cite{Bialek:2006}.\textbf{
}We comment on an approach originally suggested by Bialek and coworkers
\cite{Bialek:2006,Schneidman:2003} for revealing in how far two-point
interactions characterize the coupled dynamics of neural circuits
in retina. In fact, a systematic study of neural activity in the retina
of larval tiger salamander and guinea pig revealed that $90\%$ of
the multi-information (which measures all correlative dependencies
in a system \cite{Schneidman:2003}) is covered by pairwise correlations
only \cite{Bialek:2006}. The authors took this finding as a sign
that pairwise interactions well characterize the full network dynamics
and that higher order interactions may be neglected. Specifically,
they temporally discretized neural responses into ``1'' or ``0''
states depending on whether a neuron was or was not active during
each considered time interval of generically $20\,\mbox{ms}$. Thus,
the state of the entire (observed) network at each time interval is
given by an $N-$dimensional word composed of the binary components
of the $N$ neurons. As sometimes done for effective connectivity,
researchers often tend to go beyond what Bialek and coworkers concluded
and interpret the effective coupling strengths (\ref{eq:MaxEntropyCoupling})
as actual physical interactions of experimentally observed units.
However, it is typically not clear how correlative dependencies yield
information about direct physical interactions and such that such
interpretations are not justified without substantial further knowledge
about the system.

\subsection{Further in finding effective connectivity}

We would like to emphasize that there is a multitude of additional
approaches for finding effective or functional connectivities. For
instance, a range of methods are based on information theoretic measures
such as mutual information \cite{MacKay:2003}\textbf{, }transfer
entropy \cite{Schreiber:2000}\textbf{ }and\textbf{ }Granger causality
\cite{Granger:1969,Ladroue:2009} and extensions thereof. In addition,
there are various methods relying on Bayesian statistics or explicit
or implicit modeling or extended regression such as used in generalized
linear models \cite{Pillow:2008}. In particular in biological sciences,
such statistical methods have been used early on even at times not
many or not very reliable data were available, see for instance \cite{Aertsen:1989}
for an early study regarding effective connectivity in neural circuits.

There are a number of open challenges regarding both precision of
such methods and the interpretation of the respective results. For
instance, functional magnetic resonance imaging (fMRI) experiments
of brain areas may rely on differences in blood oxygen level (so-called
BOLD signals) as observables but often aims to relate actual dependencies
in neural spiking activity. The reasoning here is that the cell metabolism
and thus the blood oxygen consumption in a local region (typically
one cubic millimeter) of the brain is larger the more action potentials
per time are generated by the ($10^{4}-10^{6}$) neurons in that region
so that such approaches are not undisputed\textbf{ }\cite{Horwitz:2003}.
Moreover, the terms effective connectivity, functional connectivity
and structural or anatomical connectivity are sometimes not well defined,
used in different meanings across studies. There are even overlapping
definitions of non-structural forms of connectivity, e.g. for functional
connectivity, effective connectivity etc. Here we did not delve into
historic waters and used the term ``effective connectivity'' for
all forms of connectivity that is not structural. Sometimes, effective
connectivity is even taken as an indication for structural connectivity
of physical interactions. For instance, distinguishing direct interactions
from indirect influences may be an important issues (cf. Fig.~\ref{fig:StucturalVSEffectiveLinks})
that is not yet fully clarified \cite{Nawrath:2010}.

Finally, we mention that for neural circuits \cite{Cocco:2009} have
devised a statistical method to find the couplings $J_{ij}$ that
optimize the likelihood that a class of integrate and fire models
generates the spike trains observed in experiments. This statistical
method assumes the same model class as the approach for inferring
structural connectivity presented above (section \ref{sub:Reconstruction-from-spike})
and its goal indeed is finding the (most likely) structural connectivity.

As a conclusive warning, we remark that in particular different methods
exit for obtaining effective connectivity given the \emph{same} data
set; the results each provide information about specific features
of the system: exactly those defined by the method. Some might almost
coincide with others, some might be congruent with and some may well
be contradicting a given structural connectivity (cf. \cite{Luensmann:2014}).

\section{Conclusion and Open Questions\label{sec:Open-Questions}}

This review focuses on how structural connectivity of networks may
be inferred from dynamical features of the networks' nodes. It is
on purpose on an introductory level and (given that the area of network
reconstruction is rapidly growing simultaneously in different fields,
from gene and neural networks to engineering systems) by construction
only highlights some selected approaches, most of which based on a
perspective of considering the collective nonlinear network dynamics.
We provide basic approaches about finding effective or functional
connectivities of a network from time series as a brief complement
and to get a taste for essential differences in perspective. One main
distinction between approaches for identifying qualitatively different
types of connectivity is that structural inference, aiming to reconstruct
real physical interactions, take into account the time evolution of
a system. In contrast, finding effective connectivity is often based
on a stationarity assumption and uses distributions of states, neglecting
all or parts of their temporal evolution. Relating observable, possibly
effective and physical connectivity, is at the heart of current interdisciplinary
research \cite{Ren:2010,Barzel:2013}\textbf{.} As also mentioned
in the Introduction and briefly discussed in section \ref{sec:Correlation-based-methods},
functional vs. structural inference if often not well distinguished
and, in particular in early publications, the qualitative difference
in approach was sometimes not even mentioned.

The methods and approaches presented in this review aim to tell whether
or not and how strongly units in a network directly interact with
each other. This is in distinction to the entire field of system's
identification \cite{Maier:2009} where the aim is to identify the
rules underlying a dynamical system from time series and predict its
(future) dynamics based on this identification. Systems identification
for multi-dimensional systems, and thus in particular for large networks,
seems intractably hard because even if the ``real'' system is almost
(but not entirely) perfectly reconstructed, predicting their future
dynamics can be impossible due to chaos (sensitive dependence on initial
conditions), exponentially many different collective states and uncontrolled
external influences, with all these factors becoming typical for multi-dimensional
complex systems. At the same time, as in part illustrated in this
review Figure \ref{fig:Quality-of-reconstruction}, learning the existence
of strengths of interactions only (and not the precise form of dynamics)
may well be successful for much larger systems. We thus speculate
that novel methods complementing those of systems identification may
yield further insights into the interaction networks of various complex
systems.

A number of key issues are not discussed in this review but still
are sometimes pressing for progress research, in particular with respect
to applications to real world settings. We list a few:
\begin{enumerate}
\item How much information can we actually access (cf. \cite{Su:2012})?
Can we observe all the units of a network? Can we observe all dynamical
variables (dimensions) of each unit? In which sense may it make sense
to seek connectivity of nodes that are not observed?
\item What do we know a priori about the system? Are the functional forms
of network interactions known? What can we say about stability or
instability of the dynamics? Perhaps the dynamics even exhibits a
complex mixture of stable and unstable dynamics \cite{Rabinovich:2008,Ashwin:2005b,Schittler:2012}.
\item What are reasonable (or possible) number of experiments or measurements
that can be done and what is a clever trade-off against the required
computational efforts that may depend on the quality and quantity
of those measurements \cite{Tegner:2003p9754}.
\item Which of the structural features are actually the most relevant and
which are even possible. In biochemical and gene regulatory networks,
electric circuits and power grids, for instance, many interaction
links (or their absence) may have been identified by other methods
not related to network collective dynamics. Under these circumstances,
can we improve existing approaches to take such information into account?
\item How can we address genuinely stochastic dynamics, e.g. in excitable
systems, possibly even induced by small number fluctuations -- a pathway
of though that links to non-equilibrium physics, cf.~\cite{Metzner:2007}?
\item What if network connectivities change, for instance by synaptic plasticity
in neural circuits \cite{Morrison:2007,Tetzlaff:2013,Bassett:2011}
or other forms of adaptation, e.g. in social networks \cite{Skyrms:2009}.
\item ... think of your own, there are many more questions with great opportunity
for scientific progress in a range of fields ...
\end{enumerate}
Answers to all these questions will specifically restrict or enhance
the space of optional networks consistent with recorded data. From
an abstract point of view, these modify the form and dimensionality
of the set of all possible networks and points to a direction that
still addresses network dynamics as an inverse problem but goes beyond
network reconstruction. Given all restrictions, perhaps we can design
or control a network to robustly exhibit a specific functionality.
Network design and control form two additional branches in the theory
of network dynamical systems, with currently highly active research,
from biological sciences to engineering \cite{Memmesheimer:2006p10093,Memmesheimer:2006p10092,Li:2009,Liu:2011,Preciado:2012}.
Such approaches might soon be extremely influential and thought-provoking,
when, e.g., the neural and biochemical networks in our bodies as well
as our infrastructure networks surrounding us can not only be reconstructed,
but even controlled and specifically engineered. We recommend a view
point by Freeman Dyson for a very practical taste \cite{Dyson:2005}.

\,

\textbf{\emph{Acknowledgements}}\textbf{:} We thank Srinivas Gorur
Shandilya for help during project initiation and Raoul-Martin Memmesheimer,
Mor Nitzan, Dirk Witthaut, David Gross, Matthias Wendland, Sarah Hallerberg
and Hinrich Arnoldt for valuable discussions and constructive comments
on parts of the manuscript. Partially supported by the International
Max Planck Research School (IMPRS) Physics of Biological and Complex
Systems (JC) the Ministry for Education and Science (BMBF), Germany,
through the Bernstein Center for Computational Neuroscience, grant
number 01GQ1005B (MT) as well as by a grant by the Max Planck Society
to MT.

\appendix

\section{Multiple Linear Regression and $L_{2}-$Norm Minimization \label{sec:l2minimization}}

Multiple linear regression is a widely-used statistical tool for predicting
values of a set of variables depending on one independent set of variables.
Its main purpose is to infer a linear relationship between them. Thus,
the relationship between sets is assumed to have the form

\begin{equation}
\boldsymbol{y}=\boldsymbol{\text{\ensuremath{\boldsymbol{\beta}}}}x+\text{\ensuremath{\boldsymbol{\varepsilon}}},\label{eq:dyntowfixed4}
\end{equation}
where $\boldsymbol{y}\in\mathbb{R}^{1\times M}$ are the values for
a dependent variable, $\boldsymbol{\boldsymbol{\beta}}\in\mathbb{R}^{1\times K}$
the column vector of unknown linear coefficients, $x\in\mathbb{R}^{K\times M}$
is the set of values for the $K-$independent variables and $\boldsymbol{\boldsymbol{\varepsilon}}=(\varepsilon_{1},\ldots,\varepsilon_{M})\in\mathbb{R}^{1\times M}$
are random errors $\varepsilon_{i}$, $i\in\{1,\ldots,M\}$. In addition,
the errors $\varepsilon_{i}$ are assumed to be independent random
variables distributed according to a Gaussian distribution with mean
$\mu=0$ and constant variance $\sigma^{2}$ \cite{Yan:2009}.

The question is how to estimate $\boldsymbol{\beta}$ by some $\hat{\boldsymbol{\beta}}$
such that the difference between the predicted and real values, $\boldsymbol{\text{\ensuremath{\boldsymbol{\beta}}}}x$
and $\boldsymbol{y}$, is minimized? Using $L_{2}$ minimization,
the problem formally becomes

\begin{equation}
\hat{\boldsymbol{\beta}}=\arg\min_{\boldsymbol{\beta}}\|\boldsymbol{y}-\boldsymbol{\beta}x\|_{2}^{2},\label{eq:l2norm}
\end{equation}
also known as the linear least squares method \cite{Yan:2009}.

The local extremum of the $L_{2}$ norm appearing in the right hand
side of (\ref{eq:l2norm}) implies

\begin{equation}
\forall i\in\{1,\ldots,K\}\,:\,\,\,\left.\dfrac{\partial}{\partial\beta_{i}}\left[(\boldsymbol{y}-\boldsymbol{\beta}x)(\boldsymbol{y}-\boldsymbol{\beta}x)^{T}\right]\right|_{\boldsymbol{\beta}=\boldsymbol{\hat{\beta}}}=0.
\end{equation}
such that

\begin{eqnarray*}
\left.\dfrac{\partial}{\partial\beta_{i}}\left[\boldsymbol{yy}^{T}-2\boldsymbol{\beta}x\boldsymbol{y}^{T}-\boldsymbol{\beta}x(\boldsymbol{\beta}x)^{T}\right]\right|_{\boldsymbol{\beta}=\boldsymbol{\hat{\beta}}} & = & 0,\\
\Leftrightarrow\left.\dfrac{\partial\boldsymbol{\beta}}{\partial\beta_{i}}\left(x\boldsymbol{y}^{T}-x(\boldsymbol{\beta}x)^{T}\right)\right|_{\boldsymbol{\beta}=\boldsymbol{\hat{\beta}}} & = & 0,
\end{eqnarray*}
which in turn implies the best estimate

\begin{equation}
\hat{\boldsymbol{\beta}}=\boldsymbol{y}x^{T}(xx^{T})^{-1},\label{eq:l2norminimization}
\end{equation}
 of $\boldsymbol{\beta}$ according to the linear least squares method
\cite{Yan:2009}.

\section{Singular Value Decomposition and $L_{1}-$Norm Minimization\label{sec:Singular-Value-Decomposition}}

Singular Value Decomposition (SVD) is regarded as an important tool
for statisticians due to its broad variety of applications (reduced
rank regression, polar decomposition, image compression, among others
\cite{Lange:2010}). Formally, from the fundamental theorem of linear
algebra, any rectangular matrix $A\in\mathbb{R}^{m\times n}$ may
be decomposed into the product of three matrices as
\begin{equation}
A=U\Sigma V^{\mathsf{T}},\label{eq:SVD1-1}
\end{equation}
where $U\in\mathbb{R}^{m\times m}$ and $V\in\mathbb{R}^{n\times n}$
are unitary matrices with their columns referred to as \emph{left
}and \emph{right singular vectors,} respectively, and $\Sigma\in\mathbb{R}^{m\times n}$
is a rectangular diagonal matrix containing the \emph{singular values}
\cite{Strang:1993}. This decomposition is known as the \emph{Singular
Value Decomposition} of matrix $A$.

The SVD properties are useful in finding the set of all possible solutions
to an under-determined system of equations, because for every under-determined
system
\begin{equation}
A\boldsymbol{y}=\boldsymbol{b},\label{eq:SVD2-1}
\end{equation}
where $\boldsymbol{b}\in\mathbb{R}^{m\times1}$, we can use SVD (\ref{eq:SVD1-1})
to rewrite $A$ such that solving (\ref{eq:SVD2-1}) for $\boldsymbol{y}$
yields the particular solution

\begin{equation}
\boldsymbol{y}_{p}=V\tilde{\Sigma}U^{\mathsf{T}}\boldsymbol{b},\label{eq:SVD3-1}
\end{equation}
where $\tilde{\Sigma}\in\mathbb{R}^{n\times m}$ is the pseudo-inverse
of $\Sigma$ and is defined as

\begin{equation}
\tilde{\Sigma}=\Sigma^{\mathsf{T}}\left(\Sigma\Sigma^{\mathsf{T}}\right)^{-1}.\label{eq:SVD3-2}
\end{equation}

Equation (\ref{eq:SVD3-1}) defines a particular solution from the
set of all possible solutions. The general solution to equation (\ref{eq:SVD2-1})
is given by our particular solution plus a linear combination of vectors
in the null-space of $A$, i.e.,

\begin{equation}
\boldsymbol{y}=V\tilde{\Sigma}U^{\mathsf{T}}\boldsymbol{b}+V\boldsymbol{c},\label{eq:SVDfinal-1}
\end{equation}

where $\boldsymbol{c}\in\mathbb{R}^{n\times1}$ and $c_{i}=0$ for
$i\in\{1,...,r\}$ and $r=Rank(A)$. The latter ensures that the linear
combinations are made only with vectors that span the null-space.

In this case, the problem consists in maximizing the number of zero
entries in $\boldsymbol{y}$ by choosing the $c_{j}\,\,\forall\,\, j>r$
properly. This may be achieved by solving the overdetermined system
(\ref{eq:SVDfinal-1}) with $M$ equations and $M-r$ unknowns. Specifically,
it has been illustrated in \cite{Timme:2007p14319,Yeung:2002,VanBussel:2011}
that by solving 

\begin{equation}
\arg\min_{\boldsymbol{y}}\|V\tilde{\Sigma}U^{T}\boldsymbol{b}+V\boldsymbol{c}\|_{1},\label{eq:svdfinal2}
\end{equation}
employing the Barrowdale and Roberts algorithm \cite{Barrowdale:1974},
a solution having a low number (possibly the least number) of non-zero
values, consistent with the restricting equations, is often recovered.
However, there seem to not be a general proof of this observation
\cite{Boyd:2009}. In the network reconstruction context, it has been
shown that optimizing systems like (\ref{eq:svdfinal2}) (where $\boldsymbol{y}$
and $\boldsymbol{b}$ are replaced by the unit's connectivity $\boldsymbol{J}_{i}^{\mathsf{T}}$
and network constraints $\boldsymbol{Y}_{i}^{\mathsf{T}}$) yields
the actual network topology when the network is sparse (even if there
are less linear constraints than units in the network).

\section{Estimating maximum entropy parameters.}

Maximizing the entropy 

\begin{equation}
S[\rho]:=-\sum_{\boldsymbol{x}}\rho(\boldsymbol{x})\ln\rho(\boldsymbol{x}),\label{eq:ShannonEntropyAPP}
\end{equation}
we derive first the functional form of $\rho(\boldsymbol{x})$ and
second its parameters under the constraints that the non-negative
function $\rho\geq0$ is normalized 
\begin{equation}
\sum_{\boldsymbol{x}}\rho(\boldsymbol{x})=1,\label{eq:rhoNormalizationAPP}
\end{equation}
and the first moment 
\begin{equation}
\left\langle x_{i}\right\rangle _{\rho}:=\sum_{\boldsymbol{x}}x_{i}\rho(\boldsymbol{x})\overset{!}{=}\left\langle x_{i}(t)\right\rangle _{t},\label{eq:ensembleAverage1APP}
\end{equation}
 and second moment 
\begin{equation}
\left\langle x_{i}x_{j}\right\rangle _{\rho}:=\sum_{\boldsymbol{x}}x_{i}x_{j}\rho(\boldsymbol{x})\overset{!}{=}\left\langle x_{i}(t)x_{j}(t)\right\rangle _{t},\label{eq:ensembleAverage2APP}
\end{equation}
are consistent with those estimated from data of $N$ time scalar
series $x_{i}(t)$. The covariance matrix is defined via the data
as 
\begin{equation}
C_{ij}=\left\langle x_{i}(t)x_{j}(t)\right\rangle _{t}-\left\langle x_{i}(t)\right\rangle _{t}\left\langle x_{j}(t)\right\rangle _{t}\,.\label{eq:Correlationdefined}
\end{equation}
 We maximize the entropy under these $1+N+N(N-1)/2$ constraints using
a Lagrange function 
\begin{equation}
L[\rho]:=S[\rho]-a\sum_{\boldsymbol{x}}\rho(\boldsymbol{x})-\sum_{i=1}^{N}h_{i}\sum_{\boldsymbol{x}}x_{i}\rho(\boldsymbol{x})-\frac{1}{2}\sum_{i=1}^{N}\sum_{j=1}^{N}\gamma_{ij}\sum_{\boldsymbol{x}}x_{i}x_{j}\rho(\boldsymbol{x}),\label{eq:appendixC1}
\end{equation}
where we drop constants as they do not influence the location of a
maximum. Here $a$, $h_{i}$ and $\frac{1}{2}\gamma_{ij}$ are the
Lagrange multipliers to be determined. Computing the first derivatives
and equating them to zero 
\begin{equation}
-\frac{\partial L}{\partial\rho(\boldsymbol{y})}=\ln\left(\rho(\boldsymbol{y})\right)+1+a+\sum_{i=1}^{N}h_{i}y_{i}+\frac{1}{2}\sum_{i=1}^{N}\sum_{j=1}^{N}\gamma_{ij}y_{i}y_{j}\overset{!}{=}0,\label{eq:appendixc2}
\end{equation}
 yields the (unique local) maximum entropy of the form 
\begin{equation}
\begin{array}{rcl}
\rho(\boldsymbol{x}) & = & \exp\left(-1-a-\sum_{i=1}^{N}h_{i}x_{i}-\frac{1}{2}\sum_{i=1}^{N}\sum_{j=1}^{N}\gamma_{ij}x_{i}x_{j}\right)\\
 & = & \exp\left(-1-a-\boldsymbol{h}^{\textsf{T}}\boldsymbol{\boldsymbol{x}}-\frac{1}{2}\boldsymbol{x}^{\textsf{T}}\hat{J}\boldsymbol{x}\right)\\
 & = & A\exp(-\frac{1}{2}\boldsymbol{z}^{\textsf{T}}\hat{J}\boldsymbol{z})
\end{array},\label{eq:appendixc3}
\end{equation}
where we use the abbreviations $\hat{J}:=\left(\gamma_{ij}\right)_{ij}$,
$\boldsymbol{z}:=\boldsymbol{x}+\hat{J}^{-1}\boldsymbol{h}$ is an
affine function of $\boldsymbol{x}$ and 
\begin{equation}
A:=\exp\left(-1-a+\frac{1}{2}\boldsymbol{h}^{\textsf{T}}\hat{J}^{-1}\boldsymbol{h}\right),\label{eq:Adefined}
\end{equation}
 is independent of $\boldsymbol{x}$. 

In summary, this exact computation yields the probabilities of the
form
\begin{equation}
\rho(\boldsymbol{x})=Z^{-1}\exp\left(\sum_{i=1}^{N}h_{i}x_{i}+\frac{1}{2}\sum_{i=1}^{N}\sum_{j=1}^{N}\hat{J}_{ij}x_{i}x_{j}\right),\label{eq:appendixc4}
\end{equation}

where $Z^{-1}=\exp\left(-1-a\right)$.

To estimate the parameters $a$, $h_{i}$ and $\hat{J}_{ij}$ we make
the approximation that the data $\boldsymbol{x}_{i}$ form a continuous
set such that we can approximate sums by integrals. We then first
observe that 
\begin{equation}
\sum_{\boldsymbol{x}}\rho(\boldsymbol{x})\overset{\mathsf{cont}}{=}\int_{\mathbb{R}^{N}}\rho(\boldsymbol{x})d^{N}x=\int_{\mathbb{R}^{N}}Ae^{-\frac{1}{2}\boldsymbol{z}^{\textsf{T}}\hat{J}\boldsymbol{z}}d^{N}z\overset{!}{=}1,\label{eq:rhoNormalizationAPPfulfilled}
\end{equation}
due to normalization (\ref{eq:rhoNormalizationAPP}). This implies
\begin{equation}
A=(|\det(\hat{J})|/(2\pi))^{N/2},\label{eq:Acomputed}
\end{equation}
 and thus yields the parameter $a$ as a function of $\boldsymbol{h}$
and $\hat{J}$. Similarly, fixing the averages (\ref{eq:ensembleAverage1APP})
yields (approximately) 
\begin{equation}
\begin{array}{rcl}
\left\langle x_{i}\right\rangle  & = & \int_{\mathbb{R}^{N}}\, x_{i}\rho(\boldsymbol{x})d^{N}x\\
 & = & \int_{\mathbb{R}^{N}}Ae^{-\frac{1}{2}\boldsymbol{z}^{\textsf{T}}\hat{J}\boldsymbol{z}}\left(z_{i}-\sum_{j=1}^{N}\hat{(J^{-1})}_{ij}h_{j}\right)d^{N}z\\
 & = & \sum_{j=1}^{N}(\hat{J}^{-1})_{ij}h_{j}
\end{array},\label{eq:ensembleAverage1APPfulfilled}
\end{equation}
 for all $i$ and thus $\boldsymbol{h}$ as a function of $\hat{J}$.

Finally, the equations fixing the second moments 
\begin{equation}
\begin{array}{rcl}
\left\langle x_{i}x_{j}\right\rangle  & = & \int_{\mathbb{R}^{N}}\, x_{i}x_{j}\rho(\boldsymbol{x})d^{N}x\\
 & = & \int_{\mathbb{R}^{N}}x_{i}x_{j}\exp\left(-1-a-\boldsymbol{h}^{\textsf{T}}\boldsymbol{\boldsymbol{x}}-\frac{1}{2}\boldsymbol{x}^{\textsf{T}}\hat{J}\boldsymbol{x}\right)d^{N}x
\end{array},\label{eq:ensembleAverage2APPfulfilled}
\end{equation}

can be evaluated using 
\begin{equation}
F(\boldsymbol{h},\boldsymbol{x})=\exp\left(-1-a-\boldsymbol{h}^{\textsf{T}}\boldsymbol{\boldsymbol{x}}-\frac{1}{2}\boldsymbol{x}^{\textsf{T}}\hat{J}\boldsymbol{x}\right),\label{eq:Fdefined}
\end{equation}
subjected to

\begin{equation}
\frac{\partial^{2}F}{\partial h_{i}\partial h_{j}}=x_{i}x_{j}F(\boldsymbol{h},\boldsymbol{x}).\label{eq:Fderivative}
\end{equation}
 With the transformation $\boldsymbol{x}=\boldsymbol{z}-\hat{J}^{-1}\boldsymbol{h}$,
eqn. (\ref{eq:ensembleAverage2APPfulfilled}) becomes

\begin{equation}
\begin{array}{rcl}
\left\langle x_{i}x_{j}\right\rangle  & = & \frac{\partial^{2}}{\partial h_{i}\partial h_{j}}\,\int_{\mathbb{R}^{N}}F(\boldsymbol{h},\boldsymbol{x})\, d^{N}x\\
 &  & \frac{\partial^{2}}{\partial h_{i}\partial h_{j}}\,\exp(\frac{1}{2}\boldsymbol{h}^{\textsf{T}}\hat{J}^{-1}\boldsymbol{h})\,\int_{\mathbb{R}^{N}}\exp\left(-1-a-\frac{1}{2}\boldsymbol{z}^{\textsf{T}}\hat{J}\boldsymbol{z}\right)\, d^{N}z\\
 & = & \frac{\partial^{2}}{\partial h_{i}\partial h_{j}}\,\exp(\frac{1}{2}\boldsymbol{h}^{\textsf{T}}\hat{J}^{-1}\boldsymbol{h})\exp(-1-a)\left(\frac{2\pi}{\left|\det(\hat{J})\right|}\right)^{\frac{N}{2}}\\
 & = & \exp(\frac{1}{2}\boldsymbol{h}^{\textsf{T}}\hat{J}^{-1}\boldsymbol{h})\exp(-1-a)\left(\frac{2\pi}{\left|\det(\hat{J})\right|}\right)^{\frac{N}{2}}\left[\left((\hat{J}^{-1})_{i}\boldsymbol{h}\right)\left((\hat{J}^{-1})_{j}\boldsymbol{h}\right)+(\hat{J}^{-1})_{ij}\right]\\
 & = & \left\langle x_{i}\right\rangle \left\langle x_{j}\right\rangle +(\hat{J}^{-1})_{ij}
\end{array},\label{eq:appendixc5}
\end{equation}
where the last equations follows from (\ref{eq:Adefined}), (\ref{eq:Acomputed})
and (\ref{eq:ensembleAverage1APPfulfilled}).\textbf{ }Thus, using
the definition of the correlation matrix (\ref{eq:Correlationdefined})
and assuming that temporal and statistical averages are the same,
we have that the matrix 
\begin{equation}
\hat{J}=C^{-1}\label{eq:appendixc6}
\end{equation}
of effective coupling strengths between the variables is given by
the inverse of the correlation matrix of the data.

\bibliographystyle{iopart-num}
\bibliography{ReconstructionReview8}

\end{document}